\documentclass[sigconf]{acmart}

\usepackage{euscript}
\usepackage{listings}
\usepackage{cleveref}
\usepackage{algorithm}
\usepackage{algpseudocode} 
\usepackage{makecell}
\usepackage{enumerate}
\usepackage{enumitem}
\usepackage[skip=0.2pt]{subcaption}

\algrenewcommand\algorithmiccomment[1]{\hspace{2mm}\textcolor{blue}{$\triangleright$ #1}}
\algrenewcommand\algorithmicrequire{\textbf{Input:}}
\algrenewcommand\algorithmicensure{\textbf{Output:}}

\definecolor{dkgreen}{rgb}{0,0.6,0}
\definecolor{gray}{rgb}{0.5,0.5,0.5}
\definecolor{mauve}{rgb}{0.58,0,0.82}

\lstdefinestyle{sqlstyle}{
  language=SQL,
  basicstyle={\footnotesize\ttfamily},
  aboveskip=0mm,
  belowskip=1mm,
  breakatwhitespace=true,
  breaklines=true,
  classoffset=0,
  columns=flexible,
  commentstyle=\color{dkgreen},
  framexleftmargin=0.25em,
  frameshape={}{yy}{}{}, 
  keywordstyle=\color{blue},
  numbers=none, 
  numberstyle=\tiny\color{gray},
  showstringspaces=false,
  stringstyle=\color{mauve},
  tabsize=2,
  xleftmargin =1em,
  escapeinside=``
}

\newcounter{sqlcounter}

\crefname{listing}{SQL}{SQLs}

\lstnewenvironment{sqllisting}[2][]{%
  \setcounter{lstlisting}{\value{sqlcounter}}%
  \lstset{style=sqlstyle,caption={#1},label={#2}}%
}{%
  \stepcounter{sqlcounter}%
}

\newcommand{\yuxichange}[1]{{#1}}

\AtBeginDocument{%
  }

\setcopyright{acmlicensed}
\copyrightyear{2018}
\acmYear{2018}
\acmDOI{XXXXXXX.XXXXXXX}
\acmConference[Conference acronym 'XX]{Make sure to enter the correct
  conference title from your rights confirmation email}{June 03--05,
  2018}{Woodstock, NY}
\acmISBN{978-1-4503-XXXX-X/2018/06}

\newcommand{\edbtrevision}[1]{#1}
\newcommand{\vldbrevision}[1]{#1}

\let\oldsubparagraph\subparagraph
\renewcommand{\subparagraph}[1]{\smallskip\oldsubparagraph{\mbox{}\hspace*{1em}\itshape(#1)}\mbox{}}

\newcommand{\narrow}[1]{\ensuremath{\text{\scalebox{1.0}[1.0]{\textsf{#1}}}}}

\newcommand{\query}{\ensuremath{Q}}
\newcommand{\subsetq}{\ensuremath{q}} 
\newcommand{\cemodel}{\ensuremath{\mathfrak{m}}}
\newcommand{\truecard}[1]{\ensuremath{\narrow{card}(#1)}}
\newcommand{\estcard}[2][]{\ensuremath{\widehat{\smash{\narrow{card}}}_{#1}(#2)}}
\newcommand{\cards}{\ensuremath{\pmb{\kappa}}}

\newcommand{\watcheroned}{\narrow{Watcher1D}}
\newcommand{\watchertwod}{\narrow{Watcher2D}}
\newcommand{\twotier}{\narrow{SWatcher+}}

\newcommand{\true}{\narrow{True}}
\newcommand{\baseline}{\narrow{Baseline}}
\newcommand{\routinely}{\narrow{Periodic}}
\newcommand{\leo}{\narrow{LEO}}
\newcommand{\naruretrain}{\narrow{Naru-Ret}}
\newcommand{\naruinc}{\narrow{Naru-Inc}}

\newcommand{\plan}{\ensuremath{\pi}}
\newcommand{\planby}[1]{\ensuremath{\narrow{Opt}}(#1)}
\newcommand{\plancost}[1]{\ensuremath{\narrow{Cost}}(#1)}
\newcommand{\exectime}[1]{\ensuremath{\narrow{T}_\narrow{Exec}(#1)}}
\newcommand{\plantime}[1]{\ensuremath{\narrow{T}_\narrow{Opt}(#1)}}

\newcommand{\gain}{\narrow{gain}}
\newcommand{\threshgain}{\ensuremath{\theta_\gain}}
\newcommand{\threshexec}{\ensuremath{\theta_{\smash{\narrow{T}_\narrow{Exec}}}}}
\newcommand{\threshopt}{\ensuremath{\theta_{\smash{\narrow{T}_\narrow{Opt}}}}}
\newcommand{\threshwatch}{\ensuremath{\theta_{\smash{n_\narrow{watch}}}}}

\newcommand{\watchset}{\ensuremath{\EuScript{W}}}
\newcommand{\cfactors}{\ensuremath{\EuScript{F}}}
\newcommand{\similarity}{\narrow{sim}}
\newcommand{\threshsim}{\ensuremath{\theta_\similarity}}
\newcommand{\ntest}{\ensuremath{n_\narrow{test}}}
\newcommand{\ntrain}{\ensuremath{n_\narrow{train}}}
\newcommand{\nwatch}{\ensuremath{n_\narrow{watch}}}

\newcommand{\BigO}{\ensuremath{\mathrm{O}}}

\newcommand{\watcher}{\textsc{Watcher}}

\newcommand{\sql}[1]{\texttt{#1}}

\newcommand{\selection}{\sigma}
\newcommand{\join}{\Join}

\definecolor{lstpurple}{rgb}{0.5,0,0.5}
\definecolor{lstred}{rgb}{1,0,0}
\definecolor{lstreddark}{rgb}{0.7,0,0}
\definecolor{lstredl}{rgb}{0.64,0.08,0.08}
\definecolor{lstmildblue}{rgb}{0.66,0.72,0.78}
\definecolor{lstblue}{rgb}{0,0,1}
\definecolor{lstmildgreen}{rgb}{0.42,0.53,0.39}
\definecolor{lstgreen}{rgb}{0,0.5,0}
\definecolor{lstorangedark}{rgb}{0.6,0.3,0}	
\definecolor{lstorange}{rgb}{0.75,0.52,0.005}
\definecolor{lstorangelight}{rgb}{0.89,0.81,0.67}
\definecolor{lstbeige}{rgb}{0.90,0.86,0.45}

\DeclareFontShape{OT1}{cmtt}{bx}{n}{<5><6><7><8><9><10><10.95><12><14.4><17.28><20.74><24.88>cmttb10}{}

\lstdefinelanguage{smtlib2}{
  alsoletter=-,
  morekeywords={declare-const,define-fun,assert,minimize,maximize,check-sat,get-objectives,and,or,not,distinct},
  extendedchars=false,
  keywordstyle=\bfseries\color{lstpurple},
  deletekeywords={Int,Bool},
  keywords=[2]{Int,Bool},
  keywordstyle=[2]\color{lstblue},
}

\lstdefinestyle{psql}
{
tabsize=2,
basicstyle=\scriptsize\upshape\ttfamily,
language=SQL,
morekeywords={PROVENANCE,BASERELATION,INFLUENCE,COPY,ON,TRANSPROV,TRANSSQL,TRANSXML,CONTRIBUTION,COMPLETE,TRANSITIVE,NONTRANSITIVE,EXPLAIN,SQLTEXT,GRAPH,IS,ANNOT,THIS,XSLT,MAPPROV,cxpath,OF,TRANSACTION,SERIALIZABLE,COMMITTED,INSERT,INTO,WITH,SCN,UPDATED},
extendedchars=false,
keywordstyle=\bfseries,
mathescape=true,
escapechar=@,
sensitive=true
}

\lstdefinestyle{psqlcolor}
{
tabsize=2,
basicstyle=\scriptsize\upshape\ttfamily,
language=SQL,
morekeywords={PROVENANCE,BASERELATION,INFLUENCE,COPY,ON,TRANSPROV,TRANSSQL,TRANSXML,CONTRIBUTION,COMPLETE,TRANSITIVE,NONTRANSITIVE,EXPLAIN,SQLTEXT,GRAPH,IS,ANNOT,THIS,XSLT,MAPPROV,cxpath,OF,TRANSACTION,SERIALIZABLE,COMMITTED,INSERT,INTO,WITH,SCN,UPDATED},
extendedchars=false,
keywordstyle=\bfseries\color{lstpurple},
deletekeywords={count,min,max,avg,sum},
keywords=[2]{count,min,max,avg,sum},
keywordstyle=[2]\color{lstblue},
stringstyle=\color{lstreddark},
commentstyle=\color{lstgreen},
mathescape=true,
escapechar=@,
sensitive=true
}

\lstdefinestyle{datalog}
{
basicstyle=\footnotesize\upshape\ttfamily,
language=prolog
}

\lstdefinestyle{pseudocode}
{
  tabsize=3,
  basicstyle=\small,
  language=c,
  morekeywords={if,else,foreach,case,return,in,or},
  extendedchars=true,
  mathescape=true,
  literate={:=}{{$\gets$}}1 {<=}{{$\leq$}}1 {!=}{{$\neq$}}1 {append}{{$\listconcat$}}1 {calP}{{$\cal P$}}{2},
  keywordstyle=\color{lstpurple},
  escapechar=&,
  numbers=left,
  numberstyle=\color{lstgreen}\small\bfseries, 
  stepnumber=1, 
  numbersep=5pt,
}

\lstdefinestyle{xmlstyle}
{
  tabsize=3,
  basicstyle=\small,
  language=xml,
  extendedchars=true,
  mathescape=true,
  escapechar=£,
  tagstyle=\color{keywordpurple},
  usekeywordsintag=true,
  morekeywords={alias,name,id},
  keywordstyle=\color{lstred}
}

\lstdefinestyle{smtlib2}
{
tabsize=2,
basicstyle=\scriptsize\upshape\ttfamily,
numbers=left,
stepnumber=1,
breaklines=true,
stringstyle=\color{lstreddark},
commentstyle=\color{lstgreen},
mathescape=true,
escapechar=@,
sensitive=true
}

\begin{document}

\title{\twotier: A Holistic Approach to Deploying Learned Cardinality Estimators in a Dynamic Setting}

\author{Yuxi Liu}
\email{yuxi.liu@duke.edu}
\affiliation{%
    \institution{Duke University}
    \city{Durham}
    \state{NC}
    \country{USA}
}

\author{Vincent Capol}
\email{vincent.capol@duke.edu}
\affiliation{%
    \institution{Duke University}
    \city{Durham}
    \state{NC}
    \country{USA}
}

\author{Xiao Hu}
\email{xiaohu@uwaterloo.ca}
\affiliation{%
    \institution{University of Waterloo}
    \city{Waterloo}
    \state{Ontario}
    \country{Canada}
}

\author{Pankaj K. Agarwal}
\email{pankaj@cs.duke.edu}
\affiliation{%
    \institution{Duke University}
    \city{Durham}
    \state{NC}
    \country{USA}
}

\author{Jun Yang}
\email{junyang@cs.duke.edu}
\affiliation{%
    \institution{Duke University}
    \city{Durham}
    \state{NC}
    \country{USA}
}

\renewcommand{\shortauthors}{Liu et al.}

\begin{abstract}
Cardinality estimation (CE) is crucial in database query optimization.
Although recent advances have improved accuracy, deploying machine-learned CE models in dynamic databases remains challenging due to data updates and workload shifts.
Retraining can be costly, but existing practices offer little guidance on when and how to update models,
resulting in simple periodic retaining strategies that are suboptimal.
We propose {\em \twotier}, a framework that adapts learned CE to dynamic workloads by monitoring data updates and query feedback.
\vldbrevision{\twotier\ selectively monitors high-impact subqueries for which wrong estimates lead to poor plan performance,
and it proactively applies corrections to estimates during query optimization.
This strategy enables immediate adaption to data and workload shifts without waiting for full model retraining,
and carefully balances the cost and benefit of monitoring and adaption versus retraining.}
Our experiments show that \vldbrevision{\twotier{}} offers a better trade-off between query performance and overhead compared with existing approaches, making it a more practical solution for deploying learned CE in dynamic settings.
\end{abstract}

\maketitle

\section{Introduction}
\label{sec:intro}
Most database query optimizers rely heavily on \emph{cardinality estimation} (\emph{CE}) to generate query plans.
Recent advances in \emph{learned CE}~\cite{marcus2019neo,Bao,negi2020cost, hasan2020deep,marcus2018deep,park2020quicksel,10.14778/3583140.3583164, yang2019deep,repas2023selectivity,yang2020neurocard,jgmp},
where a machine learning model is trained on the database instance and/or query execution feedback,
show promise in improving CE accuracy.
However, deploying learned CE remains challenging,
because dynamic database changes can render models inadequate.
Since fine-grained incremental model updates can be expensive,
most learned models try to stay up to date by retraining or, if applicable, fine-tuning
(e.g., for deep learning~\cite{wu2021unified,yang2019deep}).
When to schedule costly refreshes, however, is unclear.
A popular approach is \emph{periodic} refresh~\cite{li2022warper,li2021cardinality,10.14778/3329772.3329780,10.14778/3476249.3476254, yang2019deep}.
The period can be defined in terms of time or the number of data updates since the last refresh,
but guidance on how to set such periods is often lacking.
There is a need to develop a smarter framework for deploying learned CE in a dynamic setting.

This paper explores several ideas to address this challenge.
The first idea is \emph{monitoring}.
We would like to monitor both data updates and query feedback for clues to refresh the model.
Ignoring data updates makes us too reactive---%
we will miss cases when data has changed significantly but not enough queries have witnessed the effect.
On the other hand, ignoring the query workload may cause unnecessary refreshes---%
often, parts of the database are rarely queried,
so refreshing the model in response to changes to these parts will not improve query performance.
The second idea is estimation \emph{correction}.
Instead of using information collected from data updates and query feedback to refresh the CE model, which can be costly,
we can use it to correct estimates by an outdated model.
This idea has been tried by LEO~\cite{LEO} years before the advent of modern learned CE models:
LEO remembers any estimation errors in excess of 5\% and uses them to correct future estimates.
There is no need to wait until the next refresh to benefit from the corrections.

However, combining these ideas into an effective solution that delivers practical, end-to-end gains is not trivial.
We approach the problem of deploying learned CE in a dynamic setting \emph{holistically}.
Instead of focusing narrowly on improving CE accuracy,
we look at the ultimate impact of model estimates on query performance,
as well as the overall operational cost, including the overhead of monitoring, refresh, and correction.
First, monitoring incurs overhead, but its benefit is not always clear.
Many CE errors may not actually impact plan quality~\cite{msrbench}.
Second, correcting CE errors for some subqueries (e.g., only single-table selections) may sometimes lead to worse plans.
This counterintuitive effect has been noted in~\cite{msrbench} and confirmed in our experiments.
Finally, retraining the CE model does not guarantee more accurate estimates where accuracy actually matters.
Overall, deploying learned CE in a dynamic setting requires a delicate balancing act between benefit and cost.

\paragraph{Contributions}

Going beyond the basic ideas of monitoring and correction, we propose \emph{\twotier}.
``+'' represents the idea of increasing the benefit of monitoring by using monitored information proactively,
to correct model estimates during optimization, without waiting for model refresh.
``S'' in \twotier\ is the idea of making monitoring and correction \emph{selective} based on their benefit, to reduce overhead.
We only monitor the cardinalities of subqueries for which correction is expected to tangibly affect plan performance.

Importantly, noting that correcting selection cardinalities alone may have little (or even negative) impact,
we also monitor information about joins.
We show that while exact single-table selection cardinalities can be maintained efficiently,
maintaining exact cardinalities of join subqueries is not cost-effective.
Instead, we only cache recent \emph{compensating correction factors} for join subqueries.
Though less accurate, such information is cheaply available even for joins involving more than two tables,
and it helps avoid the pitfall of applying only selection corrections.

Lastly, \twotier\ triggers a model refresh by detecting when it is watching too many ``correction-worthy'' cardinalities, rather than detecting a high aggregate CE error.
This policy allows \twotier\ to cap data update processing time and realize when the model is making too many consequential mistakes.

\twotier\ is also designed to be generally applicable,
without being tied to any particular learned CE model or database system.
We have experimented with using to \twotier\ to deploy three different CE models---MSCN~\cite{mscn}, JGMP~\cite{jgmp}, and Naru~\cite{yang2019deep}---with PostgreSQL, and holistically compared
its benefits and costs against competing deployment strategies.
The results demonstrate that \twotier, with a fixed configuration and minimal parameter tuning,
delivers superior end-to-end performance.
\begin{itemize}[leftmargin=*]
\item Comparing \twotier\ with the popular practice of periodic refresh,
    with MSCN as the model,
    \twotier\ achieves 44.3--101.4\% of the end-to-end running time of periodic refresh,
    outperforming it in 14 out of the 16 query templates, even when the refresh period is configured optimally post hoc.
    With JGMP and Naru, \twotier\ achieves 56.1--99.5\% and 87.84--98.84\% (resp.) of periodic refresh's end-to-end running time, beating it consistently.
\item \twotier's \emph{selective} monitoring and correction strategy readily beats
    simpler monitoring and correction strategies, including our best-effort implementation of LEO's strategy~\cite{LEO}.
\item \twotier's method of maintaining compensating correction factors for join subqueries instead of exact cardinalities
    is effective in taming the overhead while delivering query performance gains,
    even for complex queries involving multi-table joins.
\item \twotier\ naturally adapts its behavior to the workload:
    for complex OLAP-style queries, it monitors and corrects more for bigger improvements in query performance;
    for simple OLTP-style queries with little room for improvement, it automatically scales back on monitoring and correction to avoid overhead.
\end{itemize}

\section{Framework and Preliminaries}
\label{sec:prelim}

Before presenting \twotier, we first describe a general framework, called \emph{\watcher},
for deploying learned CE models in database systems in a dynamic setting;
we describe a basic monitoring strategy (\Cref{sec:watcher1d}) and 
\twotier{} (\Cref{sec:twotier}) 
within this framework.

\paragraph{Learned CE Model}

\watcher{} assumes access to a learned model $\cemodel$ that can be refreshed to predict the cardinalities of subqueries.
Model refresh in general can use query feedback---namely, queries and their true cardinalities
observed during execution---%
as well as the current database state if needed.
Our approach targets models whose refresh takes considerable time,
but we have also evaluated how it compares with, and complements, models that support incremental updates.

Importantly, \watcher{} makes no assumption about how $\cemodel$ works internally.
Our goal is not making any specific model better by modifying it;
instead, we seek to deploy a given model in the smartest way possible while working around its limitations.
Therefore, our approach is orthogonal to and complements research in improving models themselves
to be more robust in dynamic settings, e.g.,~\cite{10.14778/3583140.3583164,10.1145/3639293,wu2025practicaltheorygeneralizationselectivity}.

\paragraph{Access to Data and Queries}

\watcher{} assumes access to the data update stream (e.g., by monitoring the database log~\cite{postgre_wal}).
The framework is notified of each update and can read its content for processing.
We can also query the database for additional information as needed.
\watcher{} also assumes access to the stream of queries against the database.
Additionally, it is possible to inspect a query plan $\plan$ (for query $\query$) and its execution
to obtain the cardinality of the result set produced by each of its subplans.
With PostgreSQL, for example, such information is available with the \sql{EXPLAIN} \sql{ANALYZE} command.
However, since there are many options for $\plan$,
\watcher{} does not, in general, assume that the cardinality of every possible subquery $\subsetq$ of $\query$ is observable from the execution of $\plan$,
since $\plan$ may not contain a subplan that corresponds to $\subsetq$.
Hence, it may issue additional queries to the database to acquire subquery cardinalities;
more details are in later sections.

\paragraph{Interfacing with Query Optimizer}

\watcher{} assumes the model $\cemodel$ to be external to (as opposed to integrated into) the database query optimizer, a setup typical in learned CE research.
It can make a call $\planby{\query, \cards}$ to the optimizer
to optimize a query $\query$ using a set $\cards =\{(\subsetq_1, \estcard{\subsetq_1}), \dots, (\subsetq_k, \estcard{\subsetq_k})\}$ of cardinality estimates $\estcard{\subsetq_i}$ of some subqueries $\subsetq_i$, $1 \leq i \leq k$, of $\query$.
As is commonly done in literature~\cite{ALECE, PARQO, msrbench, han2021CEbenchmark},
instead of specifying CE for all possible subqueries, which would be too costly,
$\cards$ overrides CE only for a ``relevant'' subset of subqueries,
while leaving the estimation of others to the database optimizer (which may depend on those specified in $\cards$).
Furthermore, after obtaining a plan $\plan$,
\watcher{} asks the database system to execute the query $\query$ using it.
Finally, \twotier\ additionally assumes it can make a call $\plancost{\plan, \cards}$ to the optimizer
to estimate the cost of $\plan$ given a set $\cards$ of relevant cardinality estimates, without executing $\plan$.
For our PostgreSQL-based implementation, we use the same mechanism in~\cite{han2021CEbenchmark, PARQO} for cardinality injection,
and \sql{pg\_hint\_plan}~\cite{pghint} for specifying the plan to execute (\sql{EXPLAIN} \sql{ANALYZE}) and cost (\sql{EXPLAIN}).

\paragraph{Architecture and Overview}

\begin{figure}[t]
    \centering
    \includegraphics[width=0.92\columnwidth]{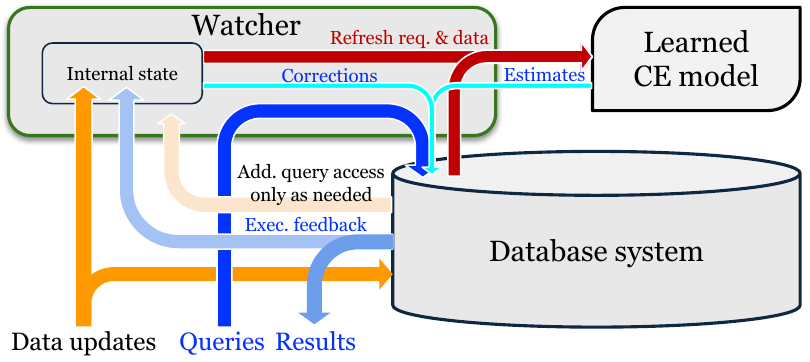}
    \vspace*{-2ex}
    \caption{\mdseries System architecture for the \watcher\ framework that interfaces with both the learned CE model and the database system.}
    \label{fig:watcher_workflow}
    \vspace*{-4ex}
\end{figure}

\Cref{fig:watcher_workflow} depicts the system architecture of \watcher{}.
The framework acts as an intermediary between the CE model and the underlying database system.
Basically, it maintains a carefully selected \emph{watchset} \watchset{} consisting of subqueries from the workload whose cardinalities are maintained up-to-date with respect to the current database state.
This practice is analogous to incrementally maintaining data summaries such as histograms in databases, but as discussed in later sections, 
we are much more careful in selecting what to maintain,
based on both the query workload and which estimation errors actually matter for query performance.
It may also maintain additional information that is updated upon data updates and/or query feedback for query optimization and triggering model refreshes. 
\watcher{}'s behavior is characterized by the following three aspects:
\begin{itemize}[leftmargin=*]
\item \emph{Processing each data update.}
    The data updates are used to incrementally update the cardinalties of \watchset{} to keep them up-to-date.
    
\item \emph{Processing each query.}
    \watcher{} intercepts each incoming query and, using the information it maintains,
    optionally corrects cardinality estimates that would otherwise be returned directly by the CE model.
    Moreover, it obtains query execution feedback (including observed true cardinalities) from the database system.
    It may update the watchset \watchset{}, i.e., insert some subqueries into \watchset{} or evict some from \watchset{}.
    It may also request additional information from the database system to update its internal state.    
\item \emph{Triggering and preparing for model refresh.}
    Using information maintained in its internal state, \watcher{} decides when to refresh the CE model.
    It leverages information already present in its internal state, and it may additionally query the database, to prepare training data for model refreshes.
\end{itemize}


We first describe a basic approach implementing the \watcher{} framework, called \emph{\watcheroned{}}, using the idea of monitoring alone (\Cref{sec:watcher1d}).
Then, we describe our main approach, called \twotier{} that maintains a more carefully selected and significantly smaller watchset and also maintains a coarse approximation of ``correction factors'' for join subqueries, thereby achieving a much better cost-benefit balance (\Cref{sec:twotier}).

\section{Basic Monitoring}
\label{sec:watcher1d}
This section describes \watcheroned, a basic implementation of our framework.
It maintains as its watchset $\watchset$
a reservoir sample of single-table selection subqueries from the query workload.
For instance, suppose we observe \Cref{sql:dsb_example} below from the query workload:
\begin{sqllisting}[\mdseries\itshape a query of TJQ013a used in \Cref{sec:experiment}]{sql:dsb_example}
SELECT * FROM store_sales, store, customer_demographics, household_demographics, customer_address, date_dim
WHERE s_store_sk = ss_store_sk
    AND ss_sold_date_sk = d_date_sk AND ss_hdemo_sk = hd_demo_sk
    AND cd_demo_sk = ss_cdemo_sk AND ss_addr_sk = ca_address_sk
    AND ss_sales_price BETWEEN 32.78 AND 207.53
    AND hd_dep_count BETWEEN 1 AND 4;
\end{sqllisting}
\watchset\ may include subqueries
$\selection_{32.78 \leq \sql{ss\_sales\_price} \leq 207.53} \sql{store\_sales}$,
and $\selection_{1 \leq \sql{hd\_dep\_count} \leq 4} \sql{household\_demographics}$.

For each $\subsetq \in \watchset$, \watcheroned\ tracks
$\estcard{\subsetq}$, the cardinality of $\subsetq$ estimated by the CE model $\cemodel$,
and $\truecard{\subsetq}$, the true cardinality of $\subsetq$.
We also index the subqueries in $\watchset$ by the tables they reference such that, for any table $R$, we can quickly retrieve $\watchset_R \subseteq \watchset$, the subset of subqueries corresponding to $R$.

Given the data update stream, \watcheroned{} tracks the true cardinalities of watchset queries,
and uses them to assess model accuracy and support model refresh.
It monitors an aggregate error measure over $\watchset$ and uses it to trigger model retraining;
the training data comes conveniently from $\watchset$, which already tracks true cardinalities.
However, our experiments show that even with careful data structure design and implementation,
the maintenance overhead can easily outweighs their potential benefit.
This limitation then motivates \emph{selective monitoring and correction} used by \twotier{} in \Cref{sec:twotier}.
Now we describe these steps in more detail:

\paragraph{Processing Queries}
For each query $\query$, we identify the set of single-table selection subqueries $\{ \subsetq \}$ as candidate watchset queries.
Whether each candidate $\subsetq$ is admitted to the watchset $\watchset$ follows the standard reservoir sampling procedure.
A previously watched query may be evicted to keep the size of $\watchset$ at $\nwatch$.

We call the optimizer with the relevant cardinalities $\hat\cards$ estimated by $\cemodel$ to obtain plan $\plan = \planby{\query, \hat\cards}$.
We record the estimated cardinality of each candidate watchset query $\subsetq$ as $\estcard{\subsetq}$.
After executing $\plan$, we record the observed true cardinality of each $\subsetq$ as $\truecard{\subsetq}$.
Such information, when available, can be parsed from the output of \sql{EXPLAIN} \sql{ANALYZE}.
Sometimes, the chosen $\plan$ may not execute $\subsetq$ as an independent subquery.\footnote{%
For example, for $\subsetq = \selection_p R$, $\plan$ may use an index-based nested-loop join where $R$ is the inner input
driven by an index scan based on a join condition in $\query$ involving $R$;
therefore, executing $\plan$ does not reveal the cardinality of $\selection_p R$
(except when the join is between the outer table's primary key and $R$'s foreign key reference thereto).}
In such cases, we execute a counting query over $\selection_p R$ to obtain $\truecard{\subsetq}$.
These additional queries are counted towards the overhead of the approach.

\paragraph{Processing Data Updates}

To process each data update concerning a row $r$ in table $R$,
for each query $\subsetq = \selection_p R$ in $\watchset_R$, we test its selection condition $p$ on $r$.
If $r$ passes $p$, we increment $\truecard{\subsetq}$ if $r$ is being inserted, or decrement $\truecard{\subsetq}$ if $r$ is being deleted.
This processing ensures that $\truecard{\cdot}$ is accurate and up to date for watchset queries.
This linear-time approach has low overhead because of its simple implementation,
which works well in practice since $\watchset_R$ is usually not large.
If $\watchset_R$ is large, more sophisticated data structures can achieve logarithmic query and update costs;
see~\cite{fullversion} for details. 

\paragraph{Retraining CE Model}

To decide when to trigger retraining of the CE model $\cemodel$,
we check an aggregate error measure, described further below, after processing each data update and query,
because both factors can affect model performance.
If either data distribution or query distribution is known to be stable,
we may decrease the frequency of checks for data updates and queries accordingly.

For the aggregate error measure, we use the $90$-th percentile of Q-error~\cite{qerror},
computed from $\truecard{\cdot}$ and $\estcard{\cdot}$ over $\ntest$ queries sampled from $\watchset$.
Unlike average or root-mean-square error, this quantile-based error is less sensitive to a handful of outliers.
Such outliers arise often in predictions of learned CE models, but they may not indicate overall model inadequacy;
a quantile-based error threshold avoids unnecessary retraining caused by these outliers.

When the aggregate error exceeds a prescribed threshold $\epsilon$,
we trigger retraining of the CE model $\cemodel$.
The training dataset is constructed by sampling $\ntrain$ watchset queries out of $\watchset$.
Since we track their up-to-date cardinalities, we do not need to rerun these queries to acquire the training data,
which significantly reduces the end-to-end latency of retraining.
Finally, retraining does not guarantee that the aggregate error will fall below $\epsilon$.
Whenever retraining fails to reach the target, we apply an exponential backoff on $\epsilon$,
i.e., setting a new target $\epsilon$ for the next retraining by multiplying $\epsilon$ by a factor of $1.5$.
If the post-training error eventually falls back to a lower previous level, the previous level can be restored.

\paragraph{Analysis}

By monitoring true cardinalities of queries sampled from the workload,
\watcheroned\ allows retraining decisions to be informed jointly by data and query distributions,
in contrast to methods~\cite{li2022warper,li2021cardinality,10.14778/3583140.3583164} that rely on less reliable heuristics.
Also, \watcheroned\ provides up-to-date training data at no additional cost,
in contrast to methods that must run additional queries for retraining~\cite{li2022warper,10.1145/3639293}.

The cost of \watcheroned\ has three components.
First, for each update on table $R$,
updating the true cardinalities for the subset $W_R$ of the watchset takes $\BigO(|W_R|)$ time (or $\BigO(\log |W_R|)$ with an advanced data structure).
Second, for each query $\query$, beyond the inference cost for $\hat\cards$,
true cardinalities are typically obtained for free from the execution;
otherwise, they are collected via additional single-table counting queries
(such cases are rare in practice and incur negligible overhead, as shown in experiments in the full version~\cite{fullversion}).
Third, retraining incurs a cost;
computing and checking the aggregate error measure has negligible cost,
and as discussed, generating the training dataset requires no extra query cost.



However, as we will see from experiments in \Cref{sec:experiment},
the query execution time savings in practice cannot justify the high overhead of maintaining \watchset{} for \watcheroned.
For example,
for a setup with query/update rate of 1/100
and 
$\nwatch=1500$,
$62.7\%$ of \watcheroned's overhead goes to processing data updates;
$4.3\%$ goes to processing queries (entirely due to CE model inference);
the remaining $33.0\%$ goes to retraining.
Overall, \watcheroned\ incurs $116$ seconds of additional overhead, only to reduce plan execution time by $66$ seconds
compared with a baseline with no retraining.





We trace this unfavorable trade-off to two limitations.
First, \watcheroned\ uses monitored information only to trigger and support model refresh,
a severe underutilization of an expensive resource.
We should derive higher, more immediate utility from actively maintained true cardinalities instead of waiting until model refresh.
Second,
\watcheroned\ focuses only on CE errors for single-table selection subqueries,
even though CE errors for join subqueries are often more critical~\cite{job,PARQO,msrbench}.
At the same time, maintaining true cardinalties for join subqueries is even costlier than for selection subqueries
(as shown by experiments on \emph{\watchertwod{}}, an extension of \watcheroned{}, in \Cref{subsec:exp_tjq}).
Therefore, we need a more cost-effective approach for join CE than incremental maintenance.
We next present \twotier\ to address these limitations.

\section{\twotier}
\label{sec:twotier}
We now describe \twotier{}, our main algorithm.
As discussed in \Cref{sec:intro},
the ``S'' in \twotier\ stands for the idea of
selectively monitoring only selection subqueries for which CE errors indeed affect plan quality;
the ``+'' in \twotier\ stands for the idea of
proactively incorporating monitored information during query optimization, instead of waiting for model refresh.
To avoid the pitfall of correcting selection CE errors alone (an observation confirmed in \Cref{subsec:exp_backfire}),
we cache compensating correction factors for joins (instead of maintaining true join cardinalities)
and apply them together with selection cardinality corrections.

\twotier\ maintains the following two main data structures:
\begin{itemize}[leftmargin=*]
\item Watchset $\watchset$.
    Like \watcheroned, $\watchset$ is a subset of single-table selection subqueries observed in the query workload.
    However, 
    instead of maintaining $\watchset$ as a reservoir sample and uses it in model refresh,
    \twotier\ admits/evicts queries to/from $\watchset$ based on their benefit to query plan efficiency,
    and proactively corrects the model \cemodel{}'s estimates with the true cardinalities in $\watchset$.
\item Cache $\cfactors$ of compensating correction factors.
    The entries are indexed by the signature of an executed query plan $\plan$ from which such factors were observed.
    More precisely, the signature has the form $(J, C)$,
    where: $J$ denotes the set of tables joined by $\plan$ and, as needed,
    the join predicate (when workload contains queries joining the same tables using different join conditions);
    $C$ denotes the (non-empty) set of single-table selection subqueries in $\watchset$
    whose monitored true cardinalities were used for correction when $\plan$ was obtained.
    Given a signature $(J, C)$, $\cfactors$ stores a mapping  $\cfactors_{J,C}$ from
    each join subplan 
    over tables $J' \subseteq J$ to a compensating correction factor denoted $\cfactors_{J,C}(J')$.
    Multiplying $\cfactors_{J,C}(J')$ with the CE (after correcting $C$) of that subplan
    would yield its observed result cardinality.
    \cfactors{} also records some basic information about the entry,
    namely the timestamp when it is cached, the query that created it, and that query's execution/planning times.
\end{itemize}

\paragraph{Overview of \twotier{} by three aspects.}

1)~To process a data update, \twotier\ follows the same procedure as that of \watcheroned\ in \Cref{sec:watcher1d}.
2)~The query processing aspect is more elaborate than that of \watcheroned{}. 
In the \emph{pre-execution stage}, \twotier\ consults $\watchset$ and $\cfactors$ to correct any model estimates for query optimization,
without waiting until refresh.
In the \emph{post-execution stage}, based on execution feedback,
\twotier{} updates information in $\watchset$ and $\cfactors$ if applicable,
and then
decides whether to admit/evict subqueries to/from $\watchset$, potentially with additional analysis overhead.
We detail the steps in \Cref{sec:twotier:query}.
3)~As for model refresh, \twotier\ decides its timing based on the number of watchset queries ---
a large number indicates that many CE errors are impacting query performance, so refresh may be beneficial.
Details are in \Cref{sec:twotier:retrain}.
Finally, \Cref{sec:twotier:discuss} concludes with cost analysis and discussion.
\Cref{tab:twotier_params} lists the parameters used by \twotier{} discussed below.

\subsection{Processing Queries}
\label{sec:twotier:query}

\paragraph{Pre-Execution}
Upon receiving a query $\query$, \twotier\ first uses the CE model $\cemodel$ to obtain the cardinality estimates $\hat\cards$.
\twotier\ then corrects $\hat\cards$ using $\watchset$ and $\cfactors$ as follows.
First, for each single-table selection subquery $\subsetq = \selection_p R$ of $\query$,
\twotier{} searches for a watchset query $\selection_{p'} R$ in $\watchset_R$ whose $p'$ matches $p$ either exactly or approximately.
We say that two conditions $p$ and $p'$ approximately match if 
their Jaccard similarity~\cite{jaccard}, denoted by $\similarity(p,p')$
(computed under practical assumptions),
is no less than a prescribed threshold $\threshsim$.
For example, consider range conditions $p = [l, u)$ and $p' = [l', u')$ on the same column.
If $p$ and $p'$ overlap,
we compute $\similarity(p,p')$ as $\smash{\frac{|\min(u, u') - \max(l, l')|}{|\max(u, u') - \min(l, l')|}}$ , or $0$ otherwise.
If multiple watchset queries have conditions that approximately match $p$,
\twotier{} selects the query $\tilde\subsetq$ with condition $\tilde p$ that achieves the highest similarity with $p$.
It then heuristically corrects\footnote{
This correction assumes an approximately uniform data distribution within query ranges, similar to histogram-based CE,
which does not always hold in practice.} the model estimate $\estcard{\subsetq}$ as
$\estcard{p \land \lnot \tilde p} + \similarity(p \land \tilde p, \tilde p) \cdot \truecard{\tilde\subsetq}$.
Intuitively, the first term estimates the cardinality of the portion of $p$ outside $\tilde p$,
while the second term scales the true cardinality of $\tilde p$ down to its intersection with $p$.
When $p$ and $\tilde p$ match exactly ($\similarity(p, \tilde p) = 1$),
the correction reduces to replacing $\estcard{\subsetq}$ with $\truecard{\tilde\subsetq}$.

\begin{figure}[t]
    \centering
    \includegraphics[width=\columnwidth]{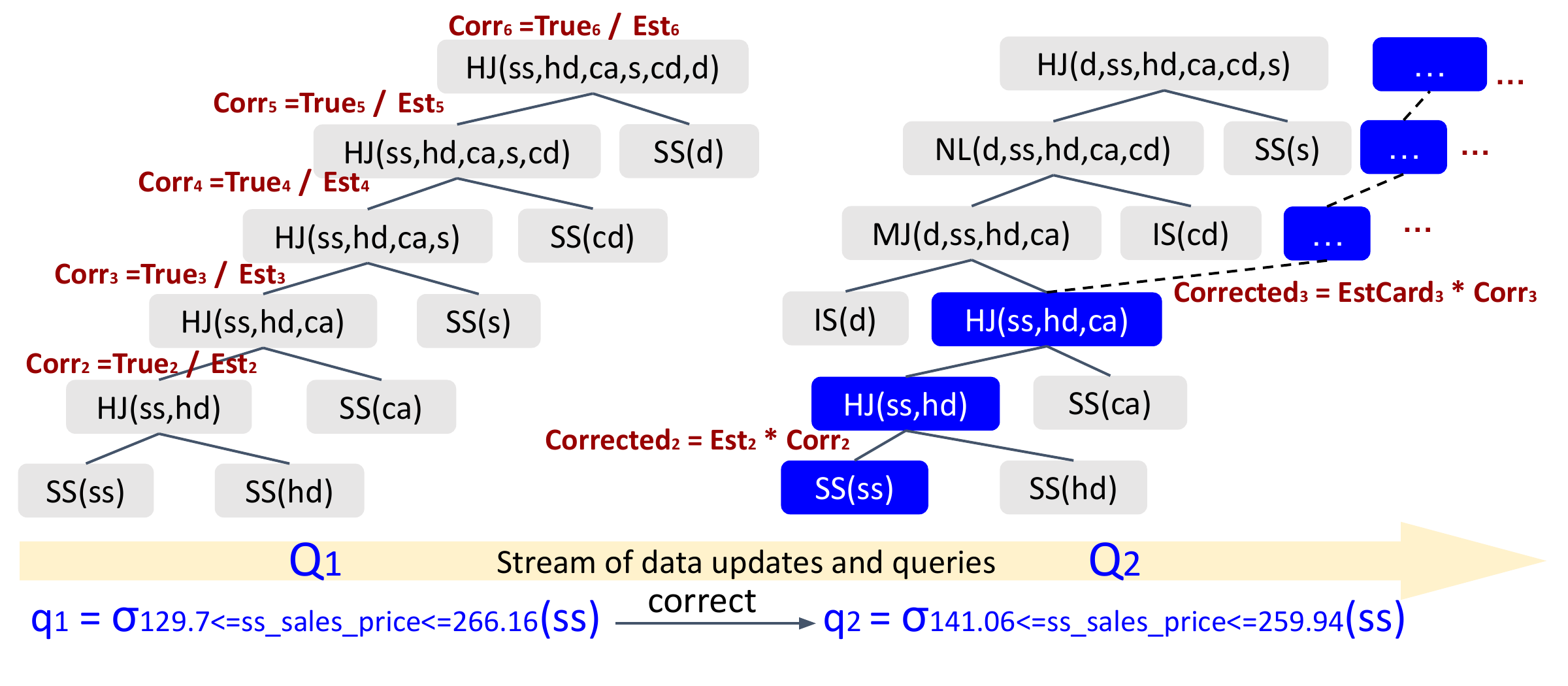}
    \vspace{-2em}
    \caption{\mdseries \twotier{}: example of corrections to both single-table selection and join cardinalities. 
    SS, HJ, IS, MJ, and NL denote the operators SeqScan, HashJoin, IndexScan, MergeJoin, and NestedLoop, respectively. 
    Corr denotes correction factors. True/Est/Corrected denote true/estimated/corrected cardinalities, respectively.}
    \label{fig:twotier_correction}
\end{figure}

\edbtrevision{
\begin{example}\label{example:twotier_correction_1}
    As shown in \Cref{fig:twotier_correction}, $\query_2$ arrives after $\query_1$ in the stream of queries and updates. Suppose that $\query_1$'s single-selection subquery $\subsetq_1$ is admitted to \watchset{} and not evicted yet. If $\subsetq_1$'s condition is the most similar to $\subsetq_2$'s --- with similarity $\frac{259.94 - 141.06}{266.16 - 129.7}$ =87.12\% exceeding the threshold $\threshsim$=80\% --- then $\truecard{\subsetq_1}$ is used to correct $\estcard{\subsetq_2}$ and $\hat\cards{[\text{\sql{ss}}]}$. The corrected CE is entered into $\tilde\cards[\text{\sql{ss}}]$.
\end{example}
}

Second, \twotier\ further corrects join cardinality estimates in $\hat\cards$ using $\cfactors$ when necessary.
Let $J_\query$ denote the set of tables joined by $\query$ and
$C_\query$ the set of watchset queries used in the correction step above.
Join CE correction may be needed if $C_\query$ is not empty.
\twotier{} first searches for \cfactors{} entries indexed by $(J_\query, C_\query)$.
If no exact match exists, it instead identifies a set of indices $\{(J_i, C_i)\}$
such that the $J_i$'s are disjoint, 
each $C_i$ equals the restriction of $C_\query$ to the tables in $J_i$,
and the union of the $J_i$'s maximally covers $J_\query$.%
\footnote{
Multiple decompositions into disjoint parts may provide a maximal cover but yield different compensating corrections.
In that case, \twotier{} prioritizes the cached query whose involved single-table subqueries maximize the minimum similarity to the current query,
favoring the closest previously observed instance.
}\textsuperscript{,}%
\footnote{This join correction procedure tries to ensure some degree of consistency when applying corrections.
Inconsistency and over-correction may arise if we correct one join cardinality using factors cached from multiple previous executions;
therefore, we require the $J_i$'s be disjoint.
Furthermore, because the observed correction factors are highly dependent on the set of corrected selections,
we require $C_i$ to be consistent with the selection corrections applied to $\query$.}
We illustrate this process in~\cite{fullversion}.
Recall that for each $(J_i, C_i)$,
$\cfactors_{J_i, C_i}$ stores the compensating correction factors for subplans joining subsets of $J_i$.
\twotier{} applies correction in a bottom-up fashion.
After correcting single-table selection CEs in the first step, consider a
join involving two tables $J'$ where at least one input cardinality is corrected.
\twotier{} first obtains the revised output CE (induced by the corrected inputs),
then multiplies it by $\cfactors_{J_i, C_i}(J')$, the compensating correction factor recorded for $J'$,
and use the result as the corrected CE.
We then repeat this process for three-table joins, four-table joins, etc.,
until the join involving all of $J_i$.

\begin{table}[t]
    \caption{\mdseries \twotier{}'s threshold parameters.}
    \label{tab:twotier_params}
    \vspace{-2ex}
    {\small
        \begin{tabular}{p{0.6cm}||c|p{5.4cm}@{ }|}
              & Stage & Usage 
             \\\hline\hline
             $\threshsim$ & Correction & Minimum similarity score for two approximately matching single-table selection conditions
             \\\hline
             $\threshexec$ & Admission & Minimum execution time for monitoring a \query{}
             \\\hline
             $\threshopt$ & Admission & Maximum optimization time for monitoring a \query{}
             \\\hline
             $\threshgain$ & Admission & Minimum estimated speed up of plan cost for monitoring a \query{}
             \\\hline
             $\threshwatch$ & Refresh & Maximum tolerated $|\watchset{}|$ before refresh
             \\\hline
        \end{tabular}}
        \vspace{-1em}
\end{table}

\begin{example}
\label{example:twotier_correction_2}
    Continuing with \Cref{example:twotier_correction_1}, $\hat\cards{[\text{\sql{ss}} \join \text{\sql{hd}}]}$ is computed from $\tilde\cards[\text{\sql{ss}}]$ and the join selectivity between \sql{ss} and \sql{hd}, estimated by the optimizer. It is refined by the compensating corrections factors in $\cfactors_{\{\text{\sql{ss}},\text{\sql{hd}},\text{\sql{ca}},\text{\sql{s}},\text{\sql{cd}},\text{\sql{d}}\}, \{\subsetq_1\}}(\text{\sql{ss}} \join \text{\sql{hd}})$, which was cached after executing $\query_1$ using the plan illustrated as the tree on the left in \Cref{fig:twotier_correction}. Then, we update $\tilde\cards{[\text{\sql{ss}} \join \text{\sql{hd}}]}$. This process is repeated for $\text{\sql{ss}} \join \text{\sql{hd}} \join \text{\sql{ca}}$, $\text{\sql{ss}} \join \text{\sql{hd}} \join \text{\sql{ca}} \join \text{\sql{s}}$, and beyond, using the compensating corrections factor stored in the index $(J_i = \{\text{\sql{ss}},\text{\sql{hd}},\text{\sql{ca}},\text{\sql{s}},\text{\sql{cd}},\text{\sql{d}}\}, C_i = \{\subsetq_1\})$. At the end, $\tilde\cards{}$, corresponding to the blue bottom-up path in the tree on the right, is injected to get the optimal plan $\planby{\query_2, \tilde\cards{}}$. Notably, as shown in \Cref{fig:twotier_correction}, it is not necessarily the same as the plan for $\query_1$.
\end{example}

Let $\tilde\cards$ be the final set of cardinality estimates with all corrections (if any),
in contrast to the set $\hat\cards$ of uncorrected cardinality estimates.
\twotier\ invokes $\planby{\query, \tilde\cards}$ to obtain the execution plan $\plan$,
and records its estimated cost $\plancost{\plan, \tilde\cards}$ and planning time $\plantime{\query}$.

\paragraph{Post-Execution} After observing the execution of $\plan$, \twotier\ first performs basic bookkeepings.
Let $\exectime{\plan}$ denote the total execution time.
If any correction was applied, \twotier{} updates the cache $\cfactors$ of compensating correction factors by creating a new mapping $\cfactors_{J_\query, C_\query}$, replacing any existing entry with the same signature.
Specifically, for each subplan of $\plan$ joining tables $J' \subseteq J_\query$,
\twotier{} records the factor $\cfactors_{J_\query, C_\query}(J')$
as the observed output cardinality of that
join subplan divided by the estimate obtained using the selection corrections in $C_\query$ but without join corrections otherwise.
Equivalently, applying this factor to the inputs corrected by \watchset{} recovers the currently observed join cardinality. 
\twotier{} also records metadata for the entry,
including the creation timestamp, query $\query$, and its observed execution/planning times.

Next, for each single-table selection subquery $\subsetq = \selection_p R$ of $\query$,
\twotier\ decides whether to admit $\subsetq$ as a watchset query (if $\subsetq \not\in \watchset$) or evict it (if $\subsetq \in \watchset$).
Conceptually, $\subsetq$ is worth monitoring if correcting its CE by its true cardinality leads to a faster query plan
by a large margin --- at least a speedup of $\threshgain$.
Since $\subsetq$'s true cardinality is available from the execution (or by an additional counting query if needed, as discussed in \Cref{sec:watcher1d}),
one could evaluate this admission condition by re-optimizing $\query$ with this correction to $\subsetq$, executing the new plan,
and measure its speedup over the plan obtained by using no CE correction.
However, the overhead would be unacceptable.
Instead, we estimate the potential improvement without any new query execution.
To further reduce the need to reoptimize,
we heuristically apply quick checks using two thresholds to eliminate any $\query$ not worthy of monitoring:
$\threshexec$, minimum execution time for $\query$,
and $\threshopt$, maximum optimization time for $\query$.
Intuitively, there is little benefit to meddle with a query that already runs fast,
and there is too much overhead to meddle with a query whose optimization takes long.
The details are explained below.

Suppose $\subsetq$ is not a watchset query.
1)~Quick checks: If $\exectime{\plan} < \threshexec$ or $\plantime{\query} > \threshopt$, do not admit $\subsetq$.
2)~Speedup check:
Let $\hat\cards$ the set of model-estimated cardinalities.
Let $\cards'$ the set of estimates formed by
the observed true cardinalities for $\subsetq$ and all join subplans of $\plan$ involving $\subsetq$,
plus uncorrected cardinality estimates for all other single-table subplans.
We invoke the optimizer to obtain the baseline plan $\hat\plan = \planby{\query, \hat\cards}$,
as well as the plan $\plan' = \planby{\query, \cards'}$, which would be obtained by admitting $\subsetq$ (alone).
Let $\cards^\star$ the set of true cardinalities observed  freely from the execution of $\plan$.
Comparing their costs under $\cards^\star$ yields the estimated speedup:
$\gain(\subsetq) = \plancost{\hat\plan, \cards^\star} / \plancost{\plan', \cards^\star}$.
We admit $\subsetq$ if and only if $\gain(\subsetq) \ge \threshgain$.
When admitting $\subsetq$ into $\watchset$,
we record its observed true cardinality $\truecard{\subsetq}$ and the speedup $\gain(\subsetq)$.

Suppose $\subsetq$ is a watchset query. We proceed with the following steps. 
1)~Quick checks: If $\exectime{\plan} \cdot \gain(\subsetq) < \threshexec$ or $\plantime{\query} > \threshopt$, evict $\subsetq$.
Here, $\exectime{\plan} \cdot \gain(\subsetq)$ estimates the execution time that would have been observed without watching $\subsetq$, using the previously recorded speedup.
2)~Speedup check: \twotier{} computes $\gain(\subsetq)$ in the exact way described above,
and evict $\subsetq$ if and only if $\gain(\subsetq) < \threshgain$.
If $\subsetq$ is evicted, we also purge the entries in $\cfactors$ indexed by any $(J,C)$ where $\subsetq \in C$; otherwise, $\gain(\subsetq)$ is updated.

\subsection{Retraining CE Model}
\label{sec:twotier:retrain}

After admitting a new watchset query, \twotier\ checks the size of the watchset.
If $|\watchset|$ exceeds a prescribed threshold $\threshwatch$, we trigger refresh of the CE model $\cemodel$.
To obtain the training set, we first add all watchset queries with their up-to-date true cardinalities, as in \Cref{sec:watcher1d}.
Since \watchset{} is small and no longer an unbiased sample of the query workload in this case,
it is augment{ed} 
by randomly sampling queries from the workload,
so that the size of the training set reaches $\ntrain$.
We assume \twotier\ has access to some representation of the query workload to sample from, such as a query log.%
\footnote{In the worst case, \twotier\ separately maintains a reservoir sample of past queries, and since the cardinalities are not tracked (unlike \Cref{sec:watcher1d}), the overhead is negligible.}
However, since these additional queries were not monitored,
\twotier{} queries the database to acquire their true cardinalities.

After retraining $\cemodel$, \twotier\ re-initializes $\watchset$ to reflect the new $\cemodel$.
This process is nearly identical to that in \Cref{sec:twotier:query}---%
specifically when $\subsetq$ is not yet a watchset query.
The key difference is that fresh $\exectime{\plan}$ and $\plantime{\query}$ are unavailable, since data updates may have occurred or the plan may change due to the new $\cemodel$.
Therefore, for each $\subsetq$, to avoid query execution, \twotier{} instead uses the most recent cached values from \watchset{} and \cfactors{}.\footnote{If multiple \cfactors{} entries (of different $\query$) contain $\subsetq$, the most recent one is used.}
Finally, we clear $\cfactors$.

Note that re-initialization of $\watchset$ does not guarantee that $|\watchset|$ fall below $\threshwatch$ following refresh.
In the unlikely event where the threshold is immediately violated,
we apply exponential backoff on $\threshwatch$, similar to the process described in \Cref{sec:watcher1d}.

\subsection{Cost Analysis and Discussion}
\label{sec:twotier:discuss}

Like \watcheroned{}, 
the cost of \twotier\ has three components: processing data updates, processing queries, and retraining.
By being more selective in which queries to monitor, the size of \watchset{} in \twotier{} is much smaller than \watcheroned{}'s---%
typically $90\%$ less
in our experiments in \Cref{sec:experiment}.
For the cost of processing queries,
the extra overhead in making CE corrections using $\watchset$ and $\cfactors$ and bookkeeping is minimal.
However, for a query $\query$ containing $k$ single-table selection subqueries,
the speed-up check 
could result in up to $k+1$ $\planby{\cdot}$ calls (one to get the baseline plan $\hat\plan$ and one for correcting each subquery)
plus $k+1$ $\plancost{\cdot}$ calls (to estimate their costs under true cardinalities).
Luckily, in practice, quick checks---i.e., assessing whether a query will run sufficiently long without corrections and whether optimization time is short---eliminate 40\%–80\% of these calls, as observed in our experiments (\Cref{sec:experiment}).
Note that an extra counting query may be needed if the plan executed did not use a selection subquery,
but such cases are rare, as discussed in \Cref{sec:watcher1d}.
On the other hand, by proactively applying CE corrections,
\twotier\ typically improves query plans and reduce query execution times more effectively than \watcheroned{}.
However, each refresh of the CE model is more expensive in \twotier{},
because \twotier\ may need to execute additional queries to construct a large enough training set,
plus additional optimizer calls (linear in the size of the training set) to re-initialize \watchset{}.
\Cref{sec:experiment} will demonstrate the overall effectiveness of \twotier\ in comparison with other approaches.

The five parameters listed in \Cref{tab:twotier_params} are straightforward to set and require little tuning (detailed in \Cref{sec:experiment}), although we also include alternative settings as ablation studies in \Cref{sec:experiment} and in~\cite{fullversion}.
As an exception, setting of $\threshwatch$ is primarily driven by users' desire to cap the overhead in processing data updates.
We believe it is possible to extend \twotier\ to avoid this parameter altogether:
with no bound on $|\watchset|$, we would control the watchset contents
by tracking both the benefit (to query performance) and cost (incurred for data updates) of each watchset query
to ensure positive net benefit over time;
model retraining would then be forced if ``thrashing'' of $\watchset$ is observed.
We leave further investigation as future work.
\section{Experiments}\label{sec:experiment}

We have conducted extensive experiments comparing \twotier\ with other methods (codes in~\cite{codes}).
To briefly summarize our results here and in the full version~\cite{fullversion},
\twotier\ consistently achieves superior overall performance,
even when competing methods are configured with post hoc-optimal settings.
\twotier's selective monitoring and correction strategies are key to its effectiveness,
allowing it to adapt gracefully to workloads with diverse characteristics ranging from OLTP to OLAP.
In particular, for OLTP queries where ML-based CE models may not be cost-effective,
\twotier\ automatically avoids unnecessary monitoring and model refreshes,
while baseline methods incur high overhead.

\paragraph{Implementation}
We implemented the \watcher{} framework in Python~3 on top of PostgreSQL V16.2,
with cardinality injection implemented using the methods in~\cite{han2021CEbenchmark,PARQO}, and plan hinting facilitated by pg\_hint\_plan~\cite{pghint}.
Reported execution times are the median of three repetitions throu-ghout our experiments. 
All experiments are conducted on a Linux server with 8 Intel(R) Core(TM) i7-7700 CPU @ 3.60GHz processors and 1TB of disk storage. 

\paragraph{CE models tested}
To assess \watcher's ability to work with different CE models,
we choose three models: MSCN~\cite{mscn}, JGMP~\cite{jgmp} and Naru~\cite{yang2019deep}.
MSCN is known for its low complexity and training/inference overhead~\cite{han2021CEbenchmark,areweready};
here we use it as a purely query-driven model by disabling MSCN's bitmaps so that it learns only from query feedback.
JGMP is a more advanced model that incorporates data samples from each table as well as PostgresSQL's up-to-date single-table selectivity estimates.
It is shown to achieve an accuracy comparable to or better than MSCN with even fewer training samples, albeit with a higher training and inference overhead.
Naru directly learns the underlying data distribution and naturally supports incremental model updates without retraining from scratch.

\paragraph{Data and updates}
We conduct experiments using both DSB~\cite{ding2021dsb,dsbgitrepo} and IMDB/JOB-light~\cite{imdb,job,job_light_sql} benchmarks.
\yuxichange{We focus primarily on DSB, as its data and queries better match our setting (see details below), and use IMDB/JOB-light to further validate \twotier{}'s effectiveness on a real-world benchmark.}
\yuxichange{DSB} contains a large amount of timestamped data,
making it easy to construct an update workload.
We have tested DSB under scale factors of 2 and 10; results on 10 are presented in~\cite{fullversion} and show similar conclusions.
To simulate data distribution shifts,
we first generate all benchmark tables under default settings.
In the DSB schema, we identify the following fact tables with timestamp columns as ``dynamic'': 
tables \sql{ss}, \sql{cs}, \sql{ws} with \sql{*\_sold\_date\_sk};
tables \sql{sr}, \sql{cr}, \sql{wr} with \sql{*\_returned\_date\_sk},
and table \sql{inv} with \sql{inv\_date\_sk}.
To generate a dynamic workload, we use a 60-day sliding window:
as it moves forward, tuples corresponding to the oldest date are deleted, and tuples for the next date are inserted.
The initial database instance contains data with \sql{*\_date\_sk} between 2450815 and 2450874.
After applying 60 days of updates, the final database instance reflects a 60-day forward shift with \sql{*\_date\_sk} between 2450875 and 2450934, representing a completely different database state.
The dimension tables remain unchanged throughout.
The third column of \Cref{tab:query_templates} lists, for some query templates we evaluated,
the total number of data updates relevant to each query, ranging from 0.08 to 1.55 million.
\yuxichange{For IMDB, following the same procedure, only \sql{title.production\_year} is a timestamp column and is  used to stream the data. 
The initial instance contains data from 1900 to 1959; 
after applying 60 years of updates, the final instance covers 1960 to 2013.
Each workload contains a total of 2.45M data updates.
}

\paragraph{Queries}
DSB features many complex query templates.
To ensure that we can run all approaches and underlying CE models and to make the results possible to interpret,
we prepare our query templates as follows:
i)~we filter out DSB query templates that did not involve any dynamic tables or lacked selection conditions on them;
ii)~we remove templates with complex predicates such as \sql{LIKE} or disjunctions;
and iii)~we further restrict the number of selection conditions in each template to one \vldbrevision{to four} (without changing join conditions).%
\footnote{\label{footnote:limit-sels}%
The reason for this restriction is to ensure the resulting ``hardness'' of the queries.
Because how query constants are generated, more selection predicates do not make a query harder---rather, query execution time tends to decrease
because of the combination of filters, thereby diminishing the impact of CE accuracy.
We show evidence for this trend later in \Cref{subsec:exp_tjq}.
Hence, we limit the number of selection predicates to ensure that our experiments cover a wide range of table selectivities.}
\Cref{tab:query_templates} summarizes the DSB templates we use.
\vldbrevision{The query templates join 5 to 9 tables and include
12 templates (prefix ``Q'') with one additional selection condition each,
2 templates (``TJQ'') with two,
1 template (``THJQ'') with three,
and 1 template (``FJQ'') with four.}
\yuxichange{For the real workload, due to space limit, we select the five largest JOB-light templates (all three 5-table  and two 4-table joins), retaining all predicates except those on \sql{production\_year}.}
The numerical IDs are the same as the numbering in DSB\yuxichange{ and JOB-light respectively.
Full queries are in the full version~\cite{fullversion}.}

Next, we generate the query workload by instantiating query constants in the templates
using uniformly sampling from the appropriate domains.
For example, a query generated from Q101 may include the selection condition
\sql{ss\_sales\_price} \sql{/} \sql{ss\_list\_price} \sql{BETWEEN} 0.49 \sql{AND} 0.89, where 0.49 and 0.89 are randomly generated.

\begin{table}[t]
    \caption{\mdseries DSB Query templates and the associated information. The templates cover a wide range of join types and predicate types. $\dagger$: the percentiles (P50, P75) and averages (Avg.) are measured over the initial query set. $\threshexec$ is the configuration for \twotier{}. The last three columns are presented in ms.}
    \label{tab:query_templates}
    \vspace{-1em}
            \small
            \begin{tabular}{c||@{ }c@{ }|@{ }r@{ }|r|r|r|}
              Template & \# tables & \# updates & P50/P75  $\ensuremath{\narrow{T}_\narrow{Exec}}\dagger$ & Avg. $\ensuremath{\narrow{T}_\narrow{Opt}}\dagger$ & $\threshexec$
             \\\hline\hline
             Q013a & 6 & 0.30M & 459/\phantom{0}542 & 3 & 400
             \\\hline
             Q013b & 6 & 0.30M & 483/\phantom{0}740 & 3 & 400
             \\\hline
             Q018 & 6 & 0.14M & 465/\phantom{0}487 & 3 & 450
             \\\hline
             Q019 & 6 & 0.30M & 206/\phantom{0}297 & 3 & 250
             \\\hline
             Q040 & 5 & 0.14M & 2/\phantom{000}2 & 3 & 2
             \\\hline
             Q072 & 9 & 1.30M & 435/\phantom{0}665 & 16 & 700
             \\\hline
             Q085a & 8 & 0.08M & 1/\phantom{000}2 & 8 & 5
             \\\hline
             Q085b & 8 & 0.08M & 1/\phantom{000}4 & 8 & 5
             \\\hline
             Q099 & 5 & 0.14M & 54/\phantom{00}83 & 2 & 80
             \\\hline
             Q100 & 8 & 0.30M & 928/1260 & 7 & 600
             \\\hline
             Q101 & 9 & 0.41M & 38/\phantom{00}46 & 67 & 50
             \\\hline
             Q102 & 9 & 1.55M & 45/\phantom{00}47 & 95 & 50
             \\\hline
             TJQ013a & 6 & 0.30M & 80/\phantom{0}143 & 3 & 400
             \\\hline
             TJQ013b & 6 & 0.30M & 96/\phantom{0}472 & 4 & 400
             \\\hline
             \vldbrevision{THJQ100} & \vldbrevision{8} & \vldbrevision{0.30M} & \vldbrevision{134/\phantom{0}310} & \vldbrevision{10} & \vldbrevision{400}
             \\\hline
             \vldbrevision{FJQ100} & \vldbrevision{8} & \vldbrevision{0.30M} & \vldbrevision{101/\phantom{0}184} & \vldbrevision{9} & \vldbrevision{400}
             \\\hline
        \end{tabular}
        \vspace*{-3ex}
\end{table}

\paragraph{Combined workloads \edbtrevision{\& evaluation metrics}}
To obtain a combined query/update workload for testing,
we generate 3000 queries and mix them at different rates with data updates:
\begin{itemize}[leftmargin=*]
    \item 3-batch: each with 20 days of updates followed by 1000 queries.
    \item 6-batch: each with 10 days of updates followed by 500 queries;
    \item 60-batch: daily updates followed by 50 queries.
\end{itemize}
Before evaluation starts on the testing workload,
all methods are given a number of training queries, randomly generated from the same query distribution as the testing workload,
along with information obtained by executing them on the initial database instance.
Such information includes single-table true cardinalities (required by \watcheroned{}, \twotier{}),
multi-table join true cardinalities (required by \watchertwod{} and baselines),
execution time, planning time, and plan cost (required by \twotier{}),
as well as any additional information required for training the underlying CE models.


Since the overall workload performance can be difficult to interpret for a mixture of different query templates,
we choose a simpler experimental design where we evaluate each query template in a separate workload.
We refer to a 6-batch workload testing a query template by the template itself (e.g., Q013a, TJQ013b), while 3-batch and 60-batch results are presented in the full version~\cite{fullversion}.
\yuxichange{We apply the same 6-batch workload generation to IMDB/JOB-light, but with only 600 queries mixed with data updates, as these queries run substantially longer than those in DSB.}

Across templates, the above workloads cover a wide range of characteristics from OLAP to OLTP. 
Since DSB is OLAP-centric, many templates are complex and expensive (e.g., Q013a, Q013b, and Q100)
and hence stand to benefit more from accurate CEs.
On the other hand, the way we extracted templates from DSB led to a few (e.g., Q040, Q101)
whose execution times are lower than their optimization times, resulting in workloads resembling OLTP. 
Intuitively, when there is little room for improving execution times, a smart system should avoid paying maintenance and optimization costs for such templates.
We keep such templates in our workloads to study how different approaches adapt to them.


\edbtrevision{
For metrics, we use Q-error~\cite{qerror} to measure CE (in-)accuracy and define \emph{overhead} as the end-to-end running time excluding query execution, including CE model inference time and the cost of maintaining required information for each method.
}

\paragraph{Methods \& parameters setup.}
For following methods,
the training set size is 1500 for MSCN and 128 for JGMP
(128 is the smallest number tested in~\cite{jgmp} in which JGMP still outperforms MSCN).
Naru uses 10 training epochs and 2 incremental update epochs (following the recommendations in~\cite{yang2019deep,naru_codebase,areweready}). 
\begin{itemize}[leftmargin=*]
    \item {\bf \true{}}: Always inject true cardinalities for monitored subqueries, with no overhead accounting.
    While unrealstic, it allows us to evaluate the benefit and limit of knowing true cardinalities.
    \item {\bf \baseline{}}: Always use the same CE model with no updates.
    \item {\bf \routinely}: Periodically refresh the model after every $X$ days of data updates,
        where $X \in \{10, 20, 30\}$ reflects varying degrees of timeliness.
        For instance, in a 6-batch setting, $X = 10$ is ideal because it triggers refresh immediately after every batch of updates, while $X = 20$ means that the 1st, 3rd, and 5th batches are handled by models that lag behind by 10 days of updates.
    \item \edbtrevision{{\bf \leo}: We implemented the monitor-and-correct approach from LEO~\cite{LEO},
        adapting it to our setting while remaining as faithful as we could.
        Specifically, \leo{} records adjustments needed for a subquery whenever the estimate deviates from the truth by at least 5\%. 
        For an incoming query, \leo{} applies an adjustment if it matches a tracked subquery with at least 80\% similarity.
        \leo{} also updates the adjustments according to data updates.
        } 
    \item {\bf \watcheroned{}}: Deploy the model with \watcheroned, using $\nwatch=\ntest=\ntrain=1500$ for MSCN and $128$ for JGMP;
        $\epsilon$ is $90$th (P90) or $95$th (P95) percentile of the Q-errors observed during training.
    \item {\bf \twotier{}}: Deploy the model with \twotier.
        We set $\threshsim{}=0.8$, $\threshwatch=600$, and $\threshopt = 10\% \cdot \threshexec$
        (ablation studies in~\cite{fullversion}) shows that changes to them within sensible ranges have little impact on results).
        For $\threshexec$, we choose a value to separate slow and fast queries in the training set when evident (see example in~\cite{fullversion});
        otherwise, we set $\threshexec$ to around the 50th or 75th percentile.
        These settings are summarized in \Cref{tab:query_templates}.
        Next, the choice of $\threshgain$, as with the training data size, depends on the characteristic of the underlying model.
        We set $\threshgain=101\%$ for MSCN/Naru and $100.01\%$ for JGMP:
        since JGMP generally produces better estimates at higher overhead,
        it makes sense for \twotier\ to be more aggressive in monitoring and correcting with a lower $\threshgain$.
        Finally, as an ablation study, we also test a \twotier{} version without correcting join subqueries,
        though the default (with such corrections) performs better overall.    
\end{itemize} 


\subsection{MSCN with/without \watcher}\label{subsec:exp_end_to_end}

We start with \yuxichange{DSB} results of applying \watcher\ with MSCN as the underlying CE model;
results for JGMP and Naru are discussed in \Cref{subsec:exp_end_to_end_jgmp}.
We focus on workloads with a single selection condition here (i.e., those starting with ``Q'');
\vldbrevision{results for multiple selection conditions (``TJQ'', ``THJQ'', and ``FJQ'')} are discussed in \Cref{subsec:exp_tjq}.
\yuxichange{The IMDB/JOB-light results are presented in \Cref{subsec:exp_end_to_end_imdb}.}

\begin{figure*}[!t]
    \centering

    \begin{subfigure}[t]{0.66\textwidth}
        \centering
        \includegraphics[width=\linewidth]{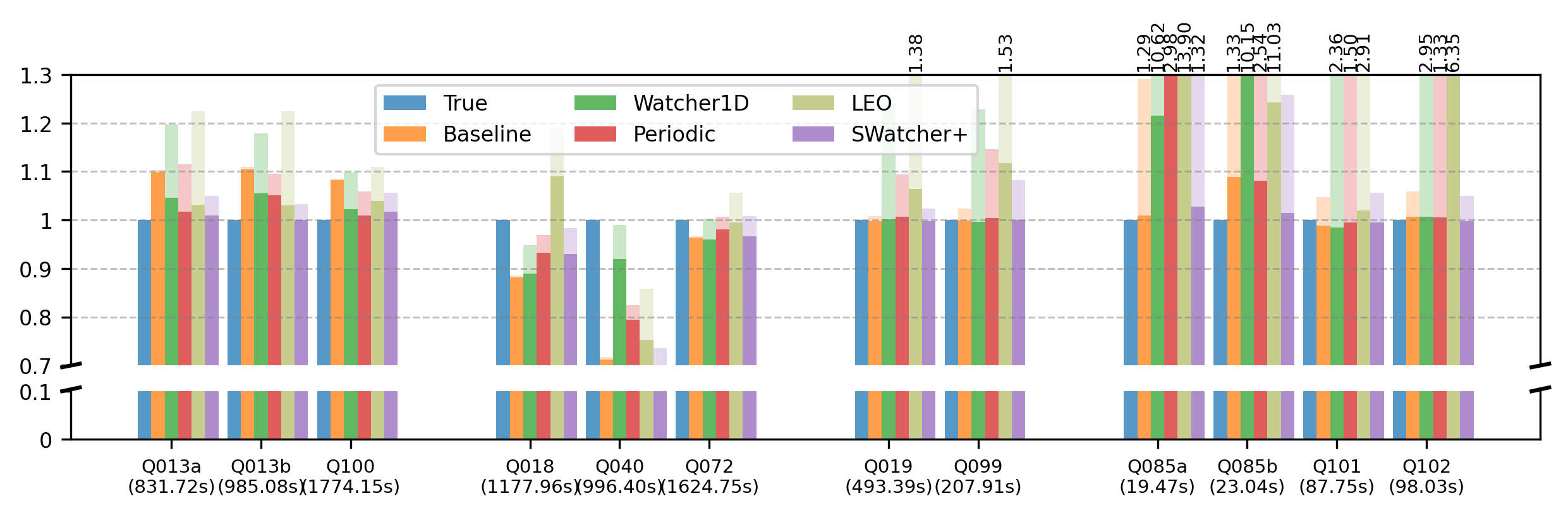}
        \caption{\mdseries Q013a, Q013b, Q100, Q018, Q040, Q072, Q019, Q099, Q085a, Q085b, Q101, Q102}
        \label{fig:exp_end_to_end}
    \end{subfigure}
    \hfill
    \begin{subfigure}[t]{0.14\textwidth}
        \centering
        \includegraphics[width=\linewidth]{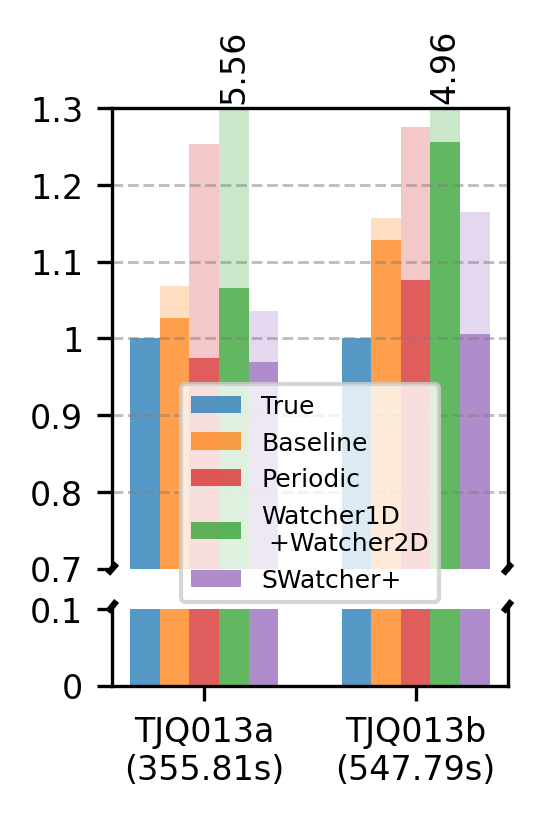}
        \caption{\mdseries TJQ013a, TJQ013b}
        \label{fig:exp_end_to_end_2d}
    \end{subfigure}
    \hfill
    \begin{subfigure}[t]{0.09\textwidth}
        \centering
        \includegraphics[width=\linewidth]{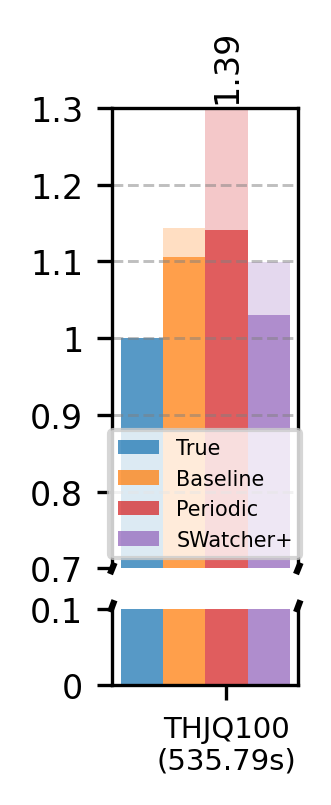}
        \caption{\mdseries THJQ100}
        \label{fig:exp_end_to_end_3d}
    \end{subfigure}
    \hfill
    \begin{subfigure}[t]{0.09\textwidth}
        \centering
        \includegraphics[width=\linewidth]{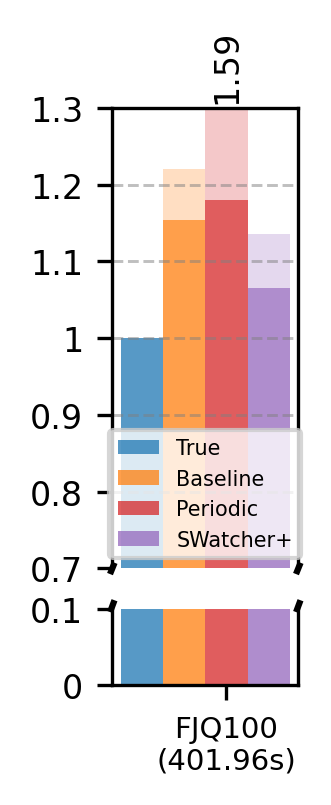}
        \caption{\mdseries FJQ100}
        \label{fig:exp_end_to_end_4d}
    \end{subfigure}
    
    \vspace{-1em}
    \caption{
    \mdseries
    \mdseries Normalized end-to-end running time, including both total query execution time (represented by the lower portion of the bar with darker shade) and overhead (upper portion with lighter shade), using MSCN as the CE model, across all methods. A number is above the bar if its height falls outside the scope. The number below the template name is the end-to-end running time of \true{}, used for normalization. \Cref{fig:exp_end_to_end} is discussed in \Cref{subsec:exp_end_to_end}, while \Cref{fig:exp_end_to_end_2d,fig:exp_end_to_end_3d,fig:exp_end_to_end_4d} are discussed in \Cref{subsec:exp_tjq}.}
    \label{fig:exp_end_to_end_mscn_and_multid}
\end{figure*}

For each single-selection query template, we measure the end-to-end running time for each method and report the results using their configuration optimized for the lowest running time.
For these workloads, we let \true\ acquire and inject true cardinalities for single-table selections for free.
We report the \routinely{} setting with the best performing period length ($X$) observed post hoc, which is impractical in a real deployment.\footnote{%
Indeed, in our experiments, \routinely{} requires different $X$ values to achieve the best trade-off between the execution time and overhead for different templates.}
We present the results for the 6-batch setting in \Cref{fig:exp_end_to_end}, and those for the 3-batch and 60-batch settings in~\cite{fullversion}.
In \Cref{fig:exp_end_to_end} (and later in \Cref{fig:exp_end_to_end_jgmp_and_naru} for JGMP and Naru), all end-to-end running times are normalized against \true{}'s, shown in the X-axis labels.
We divide the query templates into four groups below based on key observations about their results; 
we will dive deeper into Groups 1 and 2 later.

\vspace*{-1ex}
\paragraph{Group 1 (\true-improves): Q013a, Q013b, Q100.}
These templates exhibit a noticeable performance gap of approximately 10\% between \true{} and \baseline{},
indicating that injecting true selection cardinalities into the monitored subqueries significantly improves query execution times.
\twotier{} achieves the lowest end-to-end running time, outperforming \baseline{}, \routinely{}, \edbtrevision{\watcheroned{}, and \leo{}}.
By applying corrections rather than refresh, \twotier{} achieves total query execution times
that are often closer to those achieved with true cardinalities than those using updated models from \routinely{}.
In contrast, \watcheroned{} (resp.\ \leo{}) incurs much higher overhead due to the cost of maintaining a large \watchset{} (resp.\ set of adjustments),
which outweighs execution time savings.
A more detailed analysis is provided in \Cref{subsec:exp_q013b}.
   

\paragraph{Group 2 (\true-backfires): Q018, Q040, Q072.}
For this group, injecting true selection cardinalities can backfire, causing a query to execute slower than \baseline.
The reason is that PostgreSQL, while given true selection cardinalities, still relies on wrong assumptions about join selectivities
and therefore can produce worse final plans.
\twotier{} mitigates this issue by also correcting join CE using the cache $\cfactors$,\footnote{%
Another option is to use \watchertwod\ additionally to maintain true join cardinalities, but its overhead is prohibitive, as shown later in \Cref{subsec:exp_tjq}.}
while \routinely{}, \watcheroned{}, and \leo{}, which focus solely on improving selection CE accuracy, are less dependable.
Notably, \routinely{} performs best with infrequent refresh (only once), which results in lower selection CE accuracy but faster query plans than \true.
While \watcheroned{} performs okay for Q018 and Q072, its backfires poorly for Q040.
\leo{} always incurs highest overhead because it corrects selection CEs whenever possible.
Among all methods, \twotier{} is the most consistent in avoiding backfires with acceptable overhead.
\Cref{subsec:exp_backfire} provides further details.

\paragraph{Group 3 (\true-oblivious): Q019, Q099.}
This group reflects cases where query performance is relatively insensitive to CE accuracy,
resulting in similar query execution times across all methods.
Overhead cost can still make a difference:
with selective monitoring, \twotier{} achieves a 5\% to 20\% lower overhead than \routinely{} and \watcheroned{}, and much lower than \leo{}.
For Q019, \twotier{} does not add any items to its \watchset{}.
For Q099, it maintains a small \watchset{} with only 2 items, applying corrections to 1.17\% of queries,
Nonetheless, the overhead from \twotier{}'s \watchset{} admission checks still results in a slightly larger overhead than \baseline{}.
In comparison, \routinely{} and \watcheroned{} often trigger unnecessary refreshes that add more overhead without performance gains;
\leo{} incurs the highest overhead in maintaining and applying adjustments.

\paragraph{Group 4 (Opt-dominated): Q085a, Q085b, Q099, Q100.}
This group comprises queries that finish in a very short time, with total query execution times of $\leq 100$s over the entire workload.
Additionally, each query's planning time is observed to be $2\times$ to $4\times$ its execution time.
In this case, \twotier{} performs the same as \baseline{} since no queries pass the admission check, avoiding most overhead.
However, both \routinely{} and \watcheroned{} incur non-trivial overheads due to unnecessary refreshes, and \leo{} wastes resources on unnecessary adjustments.
For instance, in Q101, \routinely{} spends 39.18s for a single refresh cycle (including collecting data for retraining);
\watcheroned{} takes 20.51s for a single refresh cycle and 99.78s for maintaining \watchset{};
\leo{} spends 165.47s on adjustments.
These costs exceed the total query execution time, which remains under 90s.

\subsubsection{Analysis of \emph{\true-improves} (Q013b)}
\label{subsec:exp_q013b}

\begin{figure}[b]
\includegraphics[width=0.9\columnwidth]{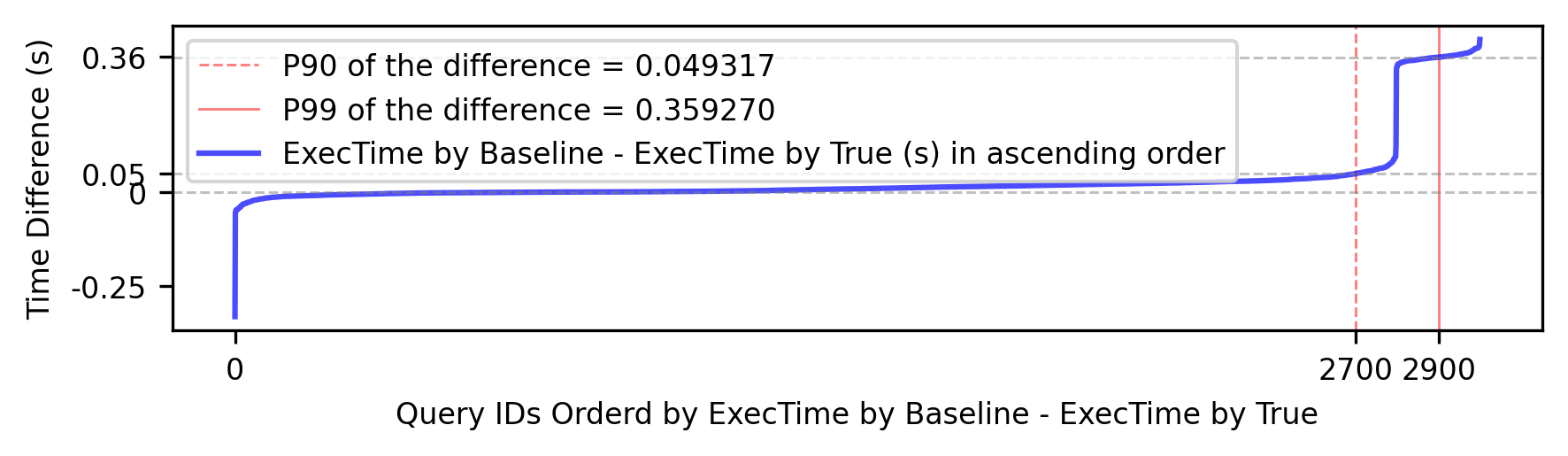}
    \vspace{-1em}
    \caption{\mdseries Q013b: ``ideal saving'' computed and ordered by the difference in execution time between \baseline{}'s and \true{}'s.}
    \label{fig:Q013b_ordered_gap}
\end{figure}

\begin{table}[t]
    \caption{\mdseries Q013b: componentized running times. $\dagger$: the parameters, shared within the same method and detailed previously, are omitted. The last five columns are in seconds. The ``Retraining Time'' column includes the number of refreshes triggered in parentheses. Highlighted rows achieve the lowest end-to-end running time (i.e., the sum of the third through sixth columns) for each method and serve as the representatives in \Cref{fig:exp_end_to_end}.}
    \label{tab:Q013b_runtime}
    \vspace{-2ex}
    {\small
        \begin{tabular}{@{ }c@{ }||@{ }r@{ }|@{ }r@{ }|@{ }r@{ }|@{ }r@{ }|@{ }r@{ }|@{ }r@{ }|}
             Method & Settings$\dagger$
             & \parbox[t]{7mm}{\rotatebox[origin=c]{90}{\makecell{Total\\ \exectime{\query}}}}
             & \parbox[t]{7mm}{\rotatebox[origin=c]{90}{\makecell{Retrainin\\ Time}}}
             & \parbox[t]{7mm}{\rotatebox[origin=c]{90}{\makecell{Overhead\\ on queries}}}
             & \parbox[t]{7mm}{\rotatebox[origin=c]{90}{\makecell{Overhead\\ on updates}}}
             & \parbox[t]{7mm}{\rotatebox[origin=c]{90}{\makecell{Total \exectime{\query}\\ in potent region}}}
             \\\hline\hline
             \textbf{\true{}} & \textbf{-} & \textbf{985.08} & \textbf{0(0)} & \textbf{0} & \textbf{0} & \textbf{75.54}
             \\\hline
             \textbf{\baseline{}} & \textbf{-} & \textbf{1087.64} & \textbf{0(0)} & \textbf{5.23} & \textbf{0} & \textbf{154.53} 
             \\\hline
             \routinely{} & X=10 & 1014.33 & 171.68(6) & 5.85 & - & 109.33
             \\\hline
             \routinely{} & X=20 & 1042.61 & 77(3) & 5.93 & - & 118.13
             \\\hline
             \textbf{\routinely{}} & \textbf{X=30} & \textbf{1035.24} & \textbf{38.04(1)} & \textbf{5.88} & \textbf{-} & \textbf{109.33}
             \\\hline
             \watcheroned{} & $\epsilon$=P90 & 1022.06 & 62.98(3) & 5.42 & 92.22 & 115.40
             \\\hline
             \textbf{\watcheroned{}} & \textbf{$\epsilon$=P95} & \textbf{1039.64} & \textbf{20.7(1)} & \textbf{5.53} & \textbf{95.37} & \textbf{128.30}
             \\\hline
             \textbf{\leo{}} & \textbf{-} & \textbf{1014.58} & \textbf{0(0)} & \textbf{19.03} & \textbf{172.67} & \textbf{82.31}
             \\\hline
             \twotier{} & \makecell[r]{$\threshgain=110\%$\\ no join \\correction} & 1021.89 & 0(0) & 28.18 & 3.70 & 110.13
             \\\hline
             \twotier{} & $\threshgain=110\%$ & 989.53 & 0(0) & 26.71 & 3.42 & 95.32 
             \\\hline
             \twotier{} & \makecell[r]{$\threshgain=101\%$\\ no join \\correction} & 1011.96 & 0(0) & 27.57 & 5.31 & 99.92
             \\\hline
             \textbf{\twotier{}} & \textbf{$\threshgain=101\%$} & \textbf{985.67} & \textbf{0(0)} & \textbf{26.31} & \textbf{4.88} & \textbf{90.68} 
             \\\hline
        \end{tabular}}
        \vspace{-1em}
\end{table}
We take a closer look at Q013b from {\bf Group 1} (\emph{\true-improves}).
First, we compare various parameter settings for all methods and show a detailed breakdown of running times in \Cref{tab:Q013b_runtime}.
For \routinely{}, refreshing after every batch of updates ($X=10$) achieves the lowest total query execution time \exectime{\query} (excluding overhead during planning)%
---over 20s shorter than the other $X$ values---though it incurs the highest retraining time and consequently loses to $X=30$ in terms of end-to-end running time.
For \watcheroned{}, using $\epsilon$ of P95 saves two refreshes (over 40s) compared with P90 but results in additional 17s of total query execution time;
both configurations incur high overhead for maintaining a \watchset{} of 1500 subqueries.
For \leo{}, applying adjustments
leads to great savings in query execution time in ``potent region'' (defined later in \Cref{fig:Q013b_ordered_gap})
but is less impressive in total \exectime{\query}.
Also, \leo{}'s overhead is prohibitive, though no retraining is needed.
Finally, for \twotier{}, starting from the baseline of $\threshgain=110\%$ and no join correction,
enabling join correction yields significant savings (>30s) in total query execution time;
lowering $\threshgain$ to $101\%$ also yields marginal improvement (<10s).
While \twotier{}'s corrections and admission-eviction checks add more overhead to queries than \routinely{}, \watcheroned{}, and \leo{},
this cost is justified by the bigger improvements in total query execution time.
Moreover, with selective monitoring, \twotier{} keeps $|\watchset{}| < 50$ in all cases,
resulting in much lower overhead in processing data updates than \watcheroned{} and \leo{}.

Excluding overhead on data updates, \watcheroned{} achieves lower running time than \routinely{} when both perform the same number of refreshes,
suggesting that \watcheroned{} identifies better timing for retraining.
\watcheroned{} also requires less retraining time than \routinely{} by skipping the collection of new training data.
Neither benefits, however, can justify \watcheroned{}'s high overhead for maintaining a large \watchset{} on data updates.

Next, as discussed in \Cref{sec:intro}, sometimes inaccurate CEs have little impact on query performance;
ideally, we would like to avoid the overhead of correcting them.
For Q013b, \Cref{fig:Q013b_ordered_gap} shows, for each of its 3000 queries,
the ``ideal saving'' achievable by \true\ over \baseline\ through injecting true selection cardinalities.
We order these queries by the amount of ideal saving.
The results show that for the leftmost 5\% queries, injection backfires;
the majority of the queries (in the middle) are insensitive to CE accuracy;
the rightmost 10\% of queries, located in what we call the ``potent region'', stand to benefit from accurate CEs,
potentially saving at least 49ms per query.
This region is exactly the target of \twotier{}.
As shown in the last column of \Cref{tab:Q013b_runtime},
\twotier{} has fast query execution time in the potent region, outperformed only by \leo{}.
However, \leo{}'s consistent adjustments strategy, agnostic to performance impact, cannot reliably produce benefits:
e.g., it has higher overall \exectime{\query} than \twotier{}, as shown in \Cref{tab:Q013b_runtime},
and it backfires more, as shown later in \Cref{subsec:exp_backfire}.
The unreliable benefit, coupled with its high overhead for processing data updates,
makes \leo{} fall short overall compared with \twotier{}.
\yuxichange{%
As a trade-off, by targeting only ``potent region'', \twotier{} does not produce CEs as accurate overall as the accuracy-driven baselines:
for the 3000 queries in Q013b, \twotier{} (across different setups) yields a P90 $\in[2.15, 2.20]$, whereas \leo{} achieves $1.20$, \routinely{} $[1.87, 2.19]$, and \watcheroned{} $[1.96, 2.19]$.
Additional percentile results are presented in the full version~\cite{fullversion}.
Nevertheless, as shown above, this accuracy gap does not always translate into a performance advantage.
}

Finally, when join correction is on and $\threshgain=101\%$,
\twotier{} selects only 1\% of single-table selection subqueries and corrects only $13\%$ of the queries in this workload.
Among queries in the potent region (\Cref{fig:Q013b_ordered_gap}), however,
\twotier{} corrects 58\% of them, and 89\% of those corrected are among the top 1\% most potent.
In short, \twotier\ focuses its efforts on queries with high payoff.



\subsubsection{Analysis of \emph{\true-backfires}}
\label{subsec:exp_backfire}
In Q018, Q040, and Q072, respectively, 16.2\%, 33\%, and 19.87\% of the queries experience a slowdown of at least 50ms per query due to injection of true cardinalities.
Meanwhile, 83\%, 66.9\%, and 75\% queries remain insensitive, and only a small fraction of queries run faster when true cardinalities are injected.
The primary reason for this backfiring is the lack of accurate estimation for other subqueries.
Correcting selection subquery cardinalities sometimes worsens join CEs, leading to a worse overall plan.
We observe serious backfiring for \leo{}, which applies adjustments to over 90\% of all queries.
Furthermore, since \leo{} is not selective about what to maintain, the overhead is significant as well.
\twotier{} mitigates the issue in two ways.
First, as observed in Q040, the admission-eviction checks effectively determine cases when injecting more accurate CEs would not improve performance, thereby avoiding caching ineffective subqueries in \watchset{} and preventing undesirable corrections.
Second, as demonstrated in Q018 and Q072, \twotier{} further applies corrections to join CEs, counteracting the inaccuracies stemmed from injecting selection subqueries solely, and guiding plan selection back on track.
Overall, among queries for which \true\ backfires,
\twotier{} achieves the following improvements in total query execution time compared with \true{}: 53s (22.3\%) in Q018, 283s (98.2\%) in Q040, and 67s (23.5\%) in Q072.
As an example, for Q018, about 89.1\% of the queries receive corrections in both selection and two-table join subqueries, contributing 73.6\% of the total savings for this query template.

The growth, utilization, and memory overhead of \watchset{} and \cfactors{}, along with the overhead of additional single-table query executions across all workloads in \Cref{fig:exp_end_to_end_mscn_and_multid}, are presented in detail in~\cite{fullversion}.

\subsection{Multi-Selections over Joins (TJQ/THJQ/FJQ)}
\label{subsec:exp_tjq}

We now turn our attention to results for templates with multiple selection conditions over joins.
We first discuss templates TJQ013a and TJQ013b, which are 6-table joins involving local selections on \sql{ss} and \sql{hd} joined on \sql{ss\_hdemo\_sk=hd\_demo\_sk}.
The three subqueries of interest are hence $\selection(\sql{ss})$, $\selection(\sql{hd})$, and $\selection(\sql{ss}) \join \selection(\sql{hd})$.
For these workloads, we let \true\ acquire and inject true cardinalities for all three subqueries of interest for free.
Here, we add the method \watcheroned{}+\watchertwod{},
where \watchertwod\ incrementally maintains cardinalities for joins involving local selections on both joining tables
(more details are available in~\cite{fullversion}).
We omit \leo{} here due to poor scalability, as its adjustment updates incur unaffordable overhead under data updates (similar to the issue faced by \watchertwod{} discussed below).
For all three \routinely{} instances, we set $X=30$, while in \watcheroned{}+\watchertwod{}, $\epsilon$ = P90.
For \baseline{}, \routinely{}, and \watcheroned{}+\watchertwod{}, the CE models are trained for all three subqueries of interest.
Finally, \twotier{} trains models for $\selection($\sql{ss}$)$ and $\selection($\sql{hd}$)$,
but relies on its \cfactors{} for join CE correction.

As shown in \Cref{fig:exp_end_to_end_2d}, for TJQ013a and TJQ013b, \twotier{} outperforms \routinely{} and \watcheroned{}+\watchertwod{},
achieving query execution times close to or, interestingly, even better than \true{}'s.
The reason is that \twotier{} goes beyond correcting the two-way join $\selection(\sql{ss}) \join \selection(\sql{hd})$;
its correction factor caching is able to capture corrections for other join patterns with acceptable overhead.
In contrast,
for \watcheroned{}+\watchertwod{}, maintaining \watchset{} in \watchertwod{} accounts for 86\% of the overhead,
exceeding 60\% of the total end-to-end running time for both templates.
This overhead renders maintaining accurate join cardinalities impractical even for two-way joins;
handling larger joins would be infeasible.

In \Cref{fig:exp_end_to_end_3d,fig:exp_end_to_end_4d}, we observe similar trends
for THJQ100 and FJQ100, both 8-table joins.
These templates have more relevant selection-join subqueries than TJQ013a and TJQ013b,
making \watcheroned{}'s monitoring prohibitively expensive;
therefore, we remove \watcheroned{}+\watchertwod{} from comparison.
Other methods are evaluated under the same setting as before.
Again, \twotier{}'s join correction using cached correction factors is lightweight and effective,
allowing it to correct join CE for subqueries such as
$\selection(\sql{ss}) \join \selection(\sql{item}) \join \selection(\sql{c})$ and
$\selection(\sql{ss}) \join \selection(\sql{item}) \join \selection(\sql{c}) \join \selection(\sql{cd})$,
without paying the cost maintaining their true cardinalities.
Even though such corrections are heuristic and approximate, they help to make \twotier{} achieve
substantially better end-to-end performance than other methods.

\begin{figure*}[!t]
    \centering
    \begin{subfigure}[t]{0.69\textwidth}
        \centering
        \includegraphics[width=\linewidth]{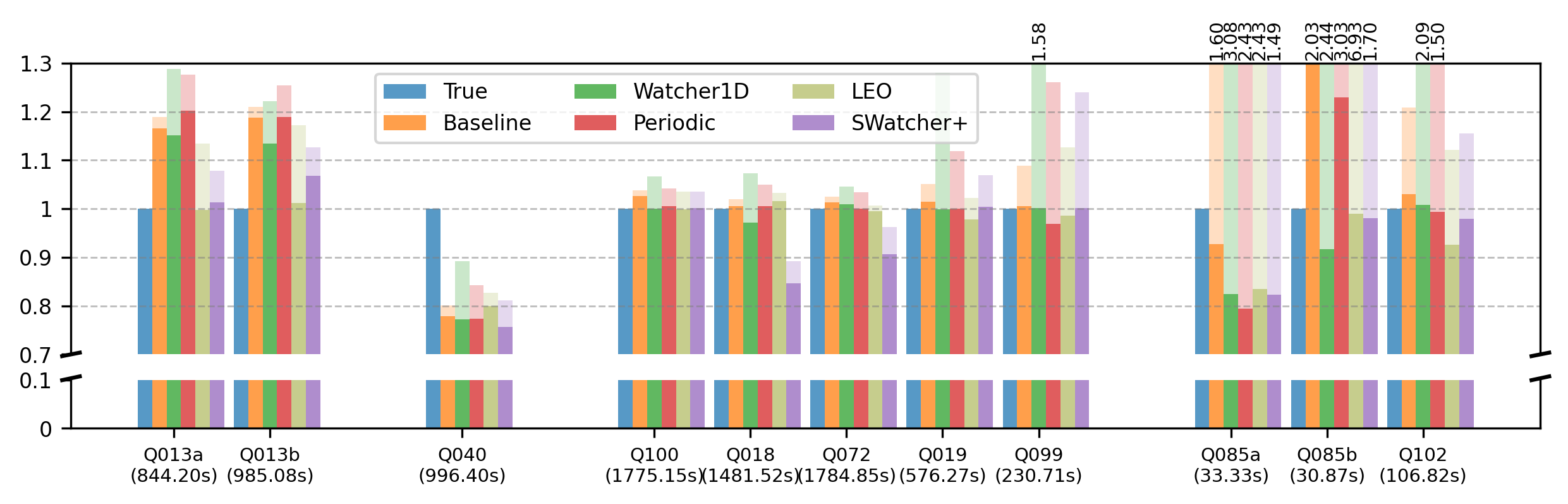}
        \caption{\mdseries \edbtrevision{
        JGMP
        }}
        \label{fig:exp_end_to_end_jgmp}
    \end{subfigure}
    \hfill
    \begin{subfigure}[t]{0.28\textwidth}
        \centering
        \includegraphics[width=\linewidth]{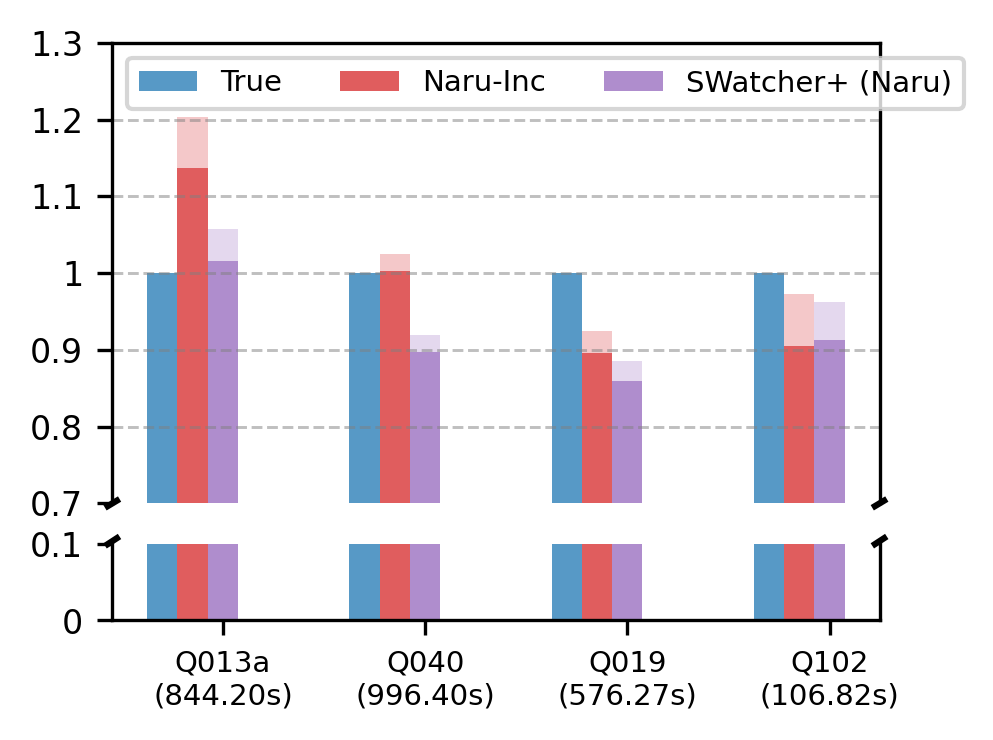}
        \caption{\mdseries 
        Naru
        }
        \label{fig:exp_end_to_end_naru}
    \end{subfigure}

    \vspace{-1em}
    \caption{
    \mdseries
    Normalized end-to-end running time using JGMP/Naru as the black-box; same format as \Cref{fig:exp_end_to_end_mscn_and_multid}.
    }
    \label{fig:exp_end_to_end_jgmp_and_naru}
\end{figure*}

Finally, as mentioned earlier at the beginning of \Cref{sec:experiment} (Footnote~\ref{footnote:limit-sels}),
we limit the number of selection predicates in query templates.
The phenomenon that additional selections tend to decrease over query execution time can be seen
here in \Cref{fig:exp_end_to_end_2d,fig:exp_end_to_end_3d,fig:exp_end_to_end_4d} when comparing TJQ013a, TJQ013b, and THJQ100/FJQ100,
and earlier in \Cref{fig:exp_end_to_end} when comparing Q013a, Q013b, and Q100.
Hence, the limit on the number of selections helps us ensure a sufficient range of table selectivities that expose more plan choices for evaluation.

\subsection{JGMP/Naru with/without \watcher}
\label{subsec:exp_end_to_end_jgmp}

In addition to MSCN, we extend the experiments (\Cref{fig:exp_end_to_end_jgmp_and_naru}) to evaluate the effectiveness of \watcher{} over a sample-efficient CE model (JGMP) and a data-driven CE model (Naru) that support incremental updates, respectively.
Despite the different design trade-offs of MSCN vs.\ JGMP vs.\ Naru,
applying \twotier\ to JGMP/Naru requires only changes to the training data size (or training/incremental update epochs) and $\threshgain$, as explained in the setup.



\paragraph{JGMP}

\Cref{fig:exp_end_to_end_jgmp} mirrors \Cref{fig:exp_end_to_end}, but shows when JGMP is used as the underlying CE model instead of MSCN.%
\footnote{The current JGMP release~\cite{jgmp_code} does not support Q101, where the selection condition is on an expression \sql{ss\_sales\_price/ss\_list\_price} rather than a single column.}
On the same workloads, JGMP achieves better CE accuracy than MSCN but incurs higher (re-)training (even with fewer training queries) costs and higher inference costs, consistent with~\cite{jgmp}.
The reason is that JGMP requires fresh samples from the tables for each retraining,
and uses PostgreSQL's up-to-date single-table selection estimates for inference,
thereby making it more aware of the current database state.
Comparing \Cref{fig:exp_end_to_end} and \Cref{fig:exp_end_to_end_jgmp}, we observe a shift:
more templates, e.g., Q100 (\emph{\true-improves} for MSCN), Q018 and Q072 (\emph{\true-backfires} for MSCN),
now fall into \emph{\true-oblivious}, indicating that JGMP's improved accuracy makes query performance less sensitive to CE errors.
It also generally reduces the adjustments \leo{} requires.
Moreover, since JGMP models are more sample-efficient, they yield a significantly smaller $\nwatch{}$ and lower overhead in \watcheroned{} compared with MSCN.
As a result, \watcheroned{} outperforms \routinely{} in two workloads---Q013b and Q085b---despite the latter's post hoc use of the optimal refresh interval.

Next, \twotier{} outperforms \baseline{}, \watcheroned{} and \routinely{} across groups \emph{\true-improves}, \emph{\true-backfires}, and \emph{opt-dominated}.
Such consistency demonstrates that \twotier{}'s corrections effectively improve query execution times and justify the overhead of maintaining \watchset{}.
Even in \emph{\true-oblivious}, which consists of workloads less sensitive to CE errors,
\twotier{} still outperforms others in 4 out of 5 templates,
because it correctly identifies the limited benefits of corrections and instead skips refreshes.
Finally, on 3 of 11 templates, Q019, Q099, and Q102, \leo{} outperforms \twotier{}.
In these cases, where CEs have little impact, JGMP's strength lets \leo{} requires fewer adjustments (though still more than \twotier{}),
while \twotier{} still incurs some overhead assessing potential performance gains before determining they are not worthwhile.
Because \twotier{} backtracks quickly, the end-to-end gap remains small and its query execution times remain competitive.

\paragraph{Naru}
Naru is a data-driven CE model that can be incrementally updated with insertions but not deletions.
Because our workloads contain both insertions and deletions, we explore two variants with day-aligned update intervals ($X$=10,20,30) following Naru's original strategy~\cite{yang2019deep}.
\naruinc{} incrementally updates the model using insertions but ignores deletions;
\naruretrain{}, retrains the model from scratch, accounting for both insertions and deletions.
Empirically, \naruinc{} performs better and is therefore selected for subsequent evaluation; additional details are in~\cite{fullversion}.

\Cref{fig:exp_end_to_end_naru} compares \naruinc{} with \twotier{} using \naruinc{} as the underlying CE model,
on one representative template for each of the four groups.
As shown in \Cref{fig:exp_end_to_end_naru}, \naruinc{}'s behavior is unstable alone,
but \twotier{}+\naruinc{} is more consistent and outperforms \naruinc{} in end-to-end running time across templates.
For Q013a (\emph{\true-improves}), \twotier{} proactively corrects the errors caused by \naruinc{} ignoring data deletions during incremental model updates.
For Q040 (\emph{\true-backfires}), \naruinc{} manages to produce reasonably accurate estimates, 
but plan performance degrades slightly due to backfiring.
For Q019 (\emph{\true-oblivious}), \naruinc{}'s inaccurate estimates happen to improve plan performance.
In all above cases, \twotier{} produces better plans than \naruinc{} with lower or comparable overhead.
Q102 (\emph{opt-dominated}) is the only template where \naruinc{}'s plans are slightly better
(though the difference is small because \emph{opt-dominated} queries are fast in the first place),
but the difference is reversed for end-to-end running time because \twotier{} incurs less overhead.
Overall, these results show that \naruinc{}, an incrementally updatable data-driven CE model, also benefits from 
\twotier{}, further demonstrate the utility and adaptivity of our framework across different CE models.

\begin{figure}[!t]
    \includegraphics[width=0.9\columnwidth]{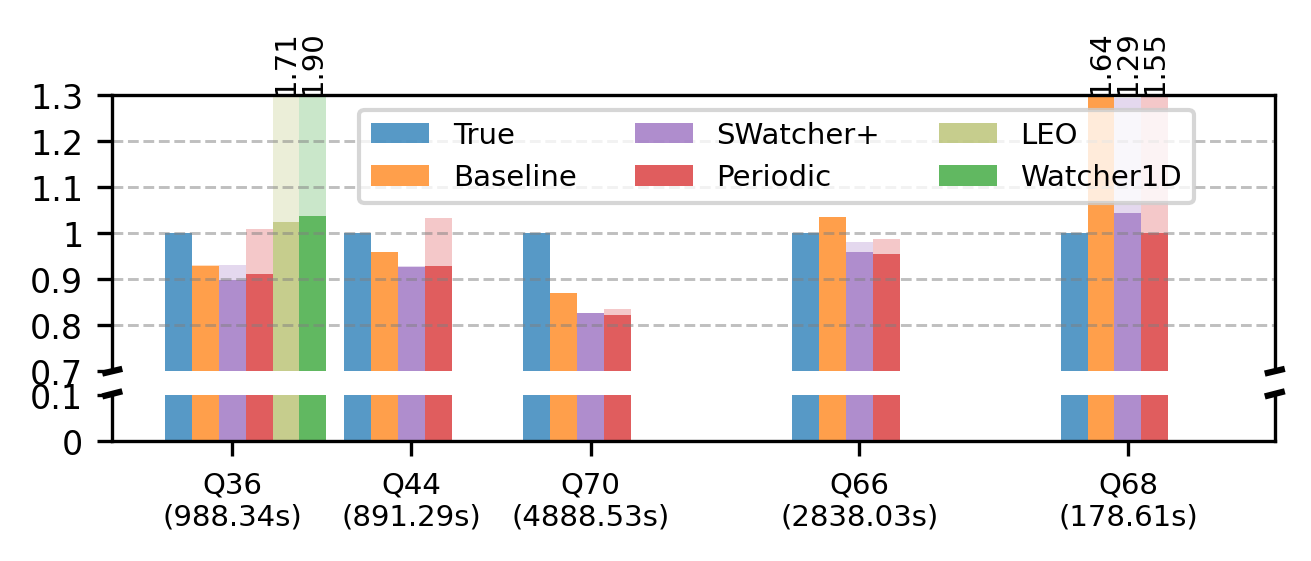}
    \vspace{-1.5em}
    \caption{\yuxichange{\mdseries Normalized end-to-end running time on IMDB using MSCN; same format as \Cref{fig:exp_end_to_end_mscn_and_multid}.}}
    \vspace{-1.5em}
    \label{fig:exp_end_to_end_imdb}
\end{figure}

\yuxichange{
\subsection{Results of IMDB/JOB-Light}
\label{subsec:exp_end_to_end_imdb}
IMDB/JOB-light results are presented in \Cref{fig:exp_end_to_end_imdb}.
We evaluate \watcheroned{} and \leo{} only on Q36, where total query execution time is roughly 1000 seconds across all methods; 
however, their maintenance overhead accounts for approximately 70\% and 90\% of this time, respectively.
Since both methods incur this maintenance overhead scaling linearly with the number of updates, which is identical across all templates, this cost is clearly prohibitive for the remaining templates as well.

We also group the five templates by observed behavior: Q36, Q44, and Q70 fall into the \emph{\true-backfires} group; Q66 into \emph{\true-oblivious}; and Q68 into \emph{opt-dominated}.
The observations are consistent with the DSB results.
Across all groups, \twotier{} consistently outperforms \routinely{}.
When accurate CEs actually backfire or have little impact, \routinely{} can perform comparably end-to-end -- but only under its least frequent retraining interval, a setting identifiable only pos hoc.
In contrast, \twotier{} adapts automatically by monitoring the impact of CE accuracy on query performance.
}

\section{Related Work}\label{sec:related}

There is growing interest in adapting learning-based CE models on evolving databases. 
Unlike \watcher{}, which is model-agnostic, these works build dedicated models that (re-)train or fine-tune upon detecting distribution shifts~\cite{areweready,10.14778/3329772.3329780,10.14778/3476249.3476254, yang2019deep,wu2021unified,park2020quicksel}. 
DDUp~\cite{10.1145/3588713} targets data drift on CE models based on neural networks. 
Warper~\cite{li2022warper} detects drift statistically and retrains, using GANs to generate new samples that are labeled by extra executions. 
Wu et al.~\cite{10.1145/3639293} mitigates workload drift with a replay buffer.
Another line of work improves model robustness:
RobustMSCN~\cite{10.14778/3583140.3583164} reduces retraining frequency at the cost of higher model complexity, and
NeuroCDF~\cite{wu2025practicaltheorygeneralizationselectivity}
enhances generalization to OOD queries.
Hybrid approaches ALECE~\cite{ALECE} and~\cite{10.1145/3654932} can directly incorporate data updates. 
Despite these advances, how to deploy CE models in dynamic environments remains unsolved.
In practice, significant data changes still necessitate frequent retraining/refreshing.

\watcher{} is a great fit for \emph{query-driven} CE methods,
due to their ability to leverage query workloads.
Recent studies on Amazon Redshift~\cite{stage} found that many real-world queries repeatedly access similar regions.
Query-driven CE models span from traditional statistical summaries~\cite{srivastava2006isomer,bruno2001stholes}
to ML-based approaches~\cite{park2020quicksel,10.14778/3329772.3329780,10.14778/3476249.3476254,wu2021unified,mscn,jgmp,10.14778/3476249.3476259,DBLP:conf/edbt/MullerWL23,10.1145/3514221.3517896}.
Among the above, we have tested \watcher{} with:
MSCN~\cite{mscn}, which uses a multi-set query representation, remains a leading query-driven cardinality estimator~\cite{areweready,10.1145/3514221.3526154}, and 
with JGMP~\cite{jgmp}, which has demonstrated better sample efficiency.

Another class of CE models is \emph{data-driven},
with underlying techniques again spanning from traditional data summaries~\cite{poosala1996improved, leis2017cardinality, lipton1990practical, poosala1997selectivity, 10.1145/3654932}) to ML~\cite{yang2019deep, wu2021unified, hilprecht2019deepdb, hasan2020deep, yang2020neurocard, zhu2020flat}.
They are generally robust to workload.
However, modeling joint data distributions is hard and less effective for complex queries,
and maintaining such models is expensive, often demanding frequent updates~\cite{10.14778/3583140.3583164}.
Specifically, DeepDB~\cite{hilprecht2019deepdb} incrementally updates the model for every insertion
(although deletion support is stated by the paper, implementation is unavailable~\cite{deepdb_codebase}).
Naru~\cite{yang2019deep} incrementally updates the model on each new batch of data appends, but batching strategy is still fixed and day-aligned.
As shown in \Cref{subsec:exp_end_to_end_jgmp}, \watcher{} can also benefit data-driven CE models,
making effective CE corrections while intelligently timing model refreshes.

The idea of guiding future optimization with query feedback has been explored for decades, beginning with LEO~\cite{LEO} and later productized in Microsoft SQL Server as CE feedback~\cite{sql_server_ce_feedback}.
The latter focuses on en/dis-abling optimizer options---such as assuming predicate correlation/independence---rather than deploying learned CE models.
LEO~\cite{LEO} targets correcting optimizer CE errors,
but as discussed in \Cref{sec:intro,sec:experiment},
its maintenance and correction decisions are based purely on CE error,
whereas \watcher{}'s decision is holistic.
Additionally, \watcher{} considers practical retraining for learned CE models,
whereas LEO assumes traditional CE models.

More recently, work on end-to-end learned query optimization~\cite{Bao,Balsa,Lero,LimeQO,BayesQO}
tackles the entire optimization problem, directly returning high-performing plans using models trained on observed executions,
rather than breaking the problem down into CE, costing, and search.
For example, BayesQO~\cite{BayesQO} uses Bayesian optimization to search for plans for recurring queries.
These approaches require substantial training and/or preparation cost, which can be amortized over time
assuming that significant data and workload distribution changes happen rarely.
Our work instead operates within the classical optimizer pipeline, focusing specifically on improving CE,
and targets a more dynamic setting.

\section{Conclusion and Future Work}
\label{sec:conclusion}

In this paper, we have studied the deployment problem of learned CE models in dynamic settings.
We take a holistic approach in both design and evaluation, considering the ultimate impact of CE on query performance,
as well as the overhead in maintaining, retraining, and using CE models.
%
%
Our results highlight the pitfall of either fixating on improving single-table CEs (simply because it is easy to do)
or overcommitting on maintaining join CEs (simply because it may yield greater improvements in query performance).
Instead, our \twotier\ effectively reduces overhead by judiciously deciding what to maintain through a benefit-based analysis.
By design, it automatically identifies high-impact CEs that warrant dedicated monitoring efforts, better handles backfiring queries caused by over-reliance on one-dimensional corrections, and avoids unnecessary retraining.
Our experiments demonstrate that \twotier\ performs well even without the benefit of hindsight, outperforming periodic retraining even when the latter is given an optimal post hoc interval.
Furthermore, \twotier\ delivers reasonable performance even in challenging cases where the benefit margin is razor-thin.
%

There are several future directions to improve \twotier,
e.g., monitoring/modeling update costs explicitly, batch-based processing of updates and counting queries, sketching for maintaining CEs under updates,
assessing benefits of CE corrections using signals beyond estimated costs, etc.,
which we detail in~\cite{fullversion}.
While there is room for improvement, we believe that \twotier{}, along with its key principles---selective monitoring, active CE correction, and a focus on end-to-end running time rather than raw CE error metrics---offers a practical solution for deploying learned CE models and valuable insights to inform future advancements in this area.



\clearpage

\bibliographystyle{ACM-Reference-Format}
\bibliography{sample}

@String{Computing = "Computing" }

@String{Computer = "{IEEE} Computer" }

@article{ding2021dsb,
  title={DSB: A decision support benchmark for workload-driven and traditional database systems},
  author={Ding, Bailu and Chaudhuri, Surajit and Gehrke, Johannes and Narasayya, Vivek},
  journal={Proceedings of the VLDB Endowment},
  volume={14},
  number={13},
  pages={3376--3388},
  year={2021},
  publisher={VLDB Endowment}
}

@article{jaccard,
    author = {Jaccard, Paul},
    year = {1901},
    month = {01},
    pages = {547-579},
    title = {Etude de la distribution florale dans une portion des Alpes et du Jura},
    volume = {37},
    journal = {Bulletin de la Societe Vaudoise des Sciences Naturelles},
    doi = {10.5169/seals-266450}
}

@article{qerror,
  title={Preventing bad plans by bounding the impact of cardinality estimation errors},
  author={Moerkotte, Guido and Neumann, Thomas and Steidl, Gabriele},
  journal={Proceedings of the VLDB Endowment},
  volume={2},
  number={1},
  pages={982--993},
  year={2009},
  publisher={VLDB Endowment}
}

@article{ALECE,
    author = {Li, Pengfei and Wei, Wenqing and Zhu, Rong and Ding, Bolin and Zhou, Jingren and Lu, Hua},
    title = {ALECE: An Attention-based Learned Cardinality Estimator for SPJ Queries on Dynamic Workloads},
    year = {2023},
    issue_date = {October 2023},
    publisher = {VLDB Endowment},
    volume = {17},
    number = {2},
    issn = {2150-8097},
    url = {https://doi.org/10.14778/3626292.3626302},
    doi = {10.14778/3626292.3626302},
    journal = {Proc. VLDB Endow.},
    month = oct,
    pages = {197–210},
    numpages = {14}
}

@article{PARQO,
  title={PARQO: Penalty-aware robust plan selection in query optimization},
  author={Xiu, Haibo and Agarwal, Pankaj K and Yang, Jun},
  journal={Proceedings of the VLDB Endowment},
  volume={17},
  number={13},
  year={2024},
  publisher={VLDB Endowment}
}

@article{msrbench,
    author = {Lee, Kukjin and Dutt, Anshuman and Narasayya, Vivek and Chaudhuri, Surajit},
    title = {Analyzing the Impact of Cardinality Estimation on Execution Plans in Microsoft SQL Server},
    year = {2023},
    issue_date = {July 2023},
    publisher = {VLDB Endowment},
    volume = {16},
    number = {11},
    issn = {2150-8097},
    url = {https://doi.org/10.14778/3611479.3611494},
    doi = {10.14778/3611479.3611494},
    journal = {Proc. VLDB Endow.},
    month = jul,
    pages = {2871–2883},
    numpages = {13}
}

@article{han2021CEbenchmark,
  title={Cardinality Estimation in DBMS: A Comprehensive Benchmark Evaluation},
  author={{Han, Yuxing and Wu, Ziniu and Wu, Peizhi and Zhu, Rong and Yang,  Jingyi and Liang, Tan Wei and Zeng, Kai and Cong, Gao and Qin, Yanzhao and Pfadler, Andreas and  Qian, Zhengping and Zhou, Jingren and Li, Jiangneng, and Cui, Bin}},
  journal={VLDB},
  year={2022},
  volume={15},
  number={4},
  issn={2150-8097}
}

@inproceedings{MSCN,
  author       = {Andreas Kipf and
                  Thomas Kipf and
                  Bernhard Radke and
                  Viktor Leis and
                  Peter A. Boncz and
                  Alfons Kemper},
  title        = {Learned Cardinalities: Estimating Correlated Joins with Deep Learning},
  booktitle    = {CIDR},
  publisher    = {www.cidrdb.org},
  year         = {2019},
  bibsource    = {dblp computer science bibliography, https://dblp.org}
}

@article{job,
    author = {Leis, Viktor and Gubichev, Andrey and Mirchev, Atanas and Boncz, Peter and Kemper, Alfons and Neumann, Thomas},
    title = {How good are query optimizers, really?},
    year = {2015},
    issue_date = {November 2015},
    publisher = {VLDB Endowment},
    volume = {9},
    number = {3},
    issn = {2150-8097},
    url = {https://doi.org/10.14778/2850583.2850594},
    doi = {10.14778/2850583.2850594},
    journal = {Proc. VLDB Endow.},
    month = nov,
    pages = {204–215},
    numpages = {12}
}

@article{areweready,
    author = {Wang, Xiaoying and Qu, Changbo and Wu, Weiyuan and Wang, Jiannan and Zhou, Qingqing},
    title = {Are we ready for learned cardinality estimation?},
    year = {2021},
    issue_date = {May 2021},
    publisher = {VLDB Endowment},
    volume = {14},
    number = {9},
    issn = {2150-8097},
    url = {https://doi.org/10.14778/3461535.3461552},
    doi = {10.14778/3461535.3461552},
    journal = {Proc. VLDB Endow.},
    month = may,
    pages = {1640–1654},
    numpages = {15}
}

@misc{pghint,
  author       = {Satoshi Nagayasu},
  title        = {pg\_hint\_plan},
  howpublished = {\url{https://github.com/ossc-db/pg_hint_plan}},
  note         = {Accessed: September 27, 2023},
  year         = 2023
}

@misc{dsbgitrepo,
  author       = {Ding, Bailu and Chaudhuri, Surajit and Gehrke, Johannes and Narasayya, Vivek},
  title        = {DSB Github Repository: SPJ templates},
  howpublished = {\url{https://github.com/microsoft/dsb/tree/main/query_templates_pg/spj_queries}},
  note         = {Accessed: September 27, 2023},
  year         = 2021
}

@inproceedings{park2020quicksel,
  title={Quicksel: Quick selectivity learning with mixture models},
  author={Park, Y. and Zhong, S. and Mozafari, B.},
  booktitle={Proc. 39th ACM SIGMOD Int. Conf. Management Data,},
  pages={1017--1033},
  year={2020}
}

@inproceedings{srivastava2006isomer,
  title={Isomer: Consistent histogram construction using query feedback},
  author={Srivastava, U. and Haas, P. J. and Markl, V. and Kutsch, M. and Tran, T. M.},
  booktitle={Proc. 22th Annu. IEEE Int. Conf. Data
Eng.},
  pages={39--39},
  year={2006}
}

@article{marcus2019neo,
  title={Neo: A learned query optimizer},
  author={Marcus, R. and Negi, P. and Mao, H. and Zhang, C. and Alizadeh, M. and Kraska, T. and Papaemmanouil, O. and Tatbul, N.},
  volume={12},
  number={11},
  pages={1705--1718},
  year={2019}
}

@inproceedings{negi2020cost,
  title={Cost-Guided Cardinality Estimation: Focus Where it Matters},
  author={Negi, P. and Marcus, R. and Mao, H. and Tatbul, N. and Kraska, T. and Alizadeh, M.},
  booktitle={ Proc. 36th Annu. IEEE Int. Conf. Data
Eng},
  pages={154--157},
  year={2020},
  organization={IEEE}
}

@inproceedings{marcus2018deep,
  title={Deep reinforcement learning for join order enumeration},
  author={Marcus, R. and Papaemmanouil, O.},
  booktitle={aiDM},
  pages={1--4},
  year={2018}
}

@inproceedings{hasan2020deep,
  title={Deep Learning Models for Selectivity Estimation of Multi-Attribute Queries},
  author={Hasan, S. and Thirumuruganathan, S. and Augustine, J. and Koudas, N. and Das, G.},
  booktitle={Proc. 39th ACM SIGMOD Int. Conf. Management Data},
  pages={1035--1050},
  year={2020}
}

@article{yang2019deep,
  title={Deep unsupervised cardinality estimation},
  author={Yang, Z. and Liang, E. and Kamsetty, A. and Wu, C. and Duan, Y. and Chen, X. and Abbeel, P. and Hellerstein, J. M. and Krishnan, S. and Stoica, I.},
  year={2019}, 
  volume = {13},
  number = {3},
  journal = {Proc. VLDB Endow.},
  pages = {279–292},
  numpages = {14}
}

@article{hilprecht2019deepdb,
  title={DeepDB: learn from data, not from queries!},
  author={Hilprecht, B. and Schmidt, A. and Kulessa, M. and Molina, A. and Kersting, K. and Binnig, C.},
  journal={Proc. VLDB Endow.},
  volume={13},
  number={7},
  pages={992--1005},
  year={2019}
}

@inproceedings{bruno2001stholes,
  title={STHoles: A multidimensional workload-aware histogram},
  author={Bruno, N. and Chaudhuri, S. and Gravano, L.},
  booktitle={Proc. 20th ACM SIGMOD Int. Conf. Management Data},
  pages={211--222},
  year={2001}
}

@article{poosala1996improved,
  title={Improved histograms for selectivity estimation of range predicates},
  author={Poosala, V. and Haas, P. J. and Ioannidis, Y. E. and Shekita, E. J.},
  journal={ACM Sigmod Record},
  volume={25},
  number={2},
  pages={294--305},
  year={1996},
  publisher={ACM New York, NY, USA}
}

@inproceedings{poosala1997selectivity,
  title={Selectivity estimation without the attribute value independence assumption},
  author={Poosala, V. and Ioannidis, Y. E.},
  booktitle={VLDB},
  volume={97},
  pages={486--495},
  year={1997},
  organization={Citeseer}
}

@inproceedings{lipton1990practical,
  title={Practical selectivity estimation through adaptive sampling},
  author={Lipton, R. J. and Naughton, J. F. and Schneider, D. A.},
  booktitle={Proc. 9th ACM SIGMOD Int. Conf. Management Data},
  pages={1--11},
  year={1990}
}

@article{repas2023selectivity,
  title={Selectivity Estimation of Inequality Joins in Databases},
  author={Repas, Diogo and Luo, Zhicheng and Schoemans, Maxime and Sakr, Mahmoud},
  journal={Mathematics},
  volume={11},
  number={6},
  pages={1383},
  year={2023},
  publisher={MDPI}
}

@article{yang2020neurocard,
  title={NeuroCard: one cardinality estimator for all tables},
  author={Yang, Zongheng and Kamsetty, Amog and Luan, Sifei and Liang, Eric and Duan, Yan and Chen, Xi and Stoica, Ion},
  journal={Proceedings of the VLDB Endowment},
  volume={14},
  number={1},
  pages={61--73},
  year={2020},
  publisher={VLDB Endowment}
}

@inproceedings{leis2017cardinality,
  title={Cardinality Estimation Done Right: Index-Based Join Sampling.},
  author={Leis, Viktor and Radke, Bernhard and Gubichev, Andrey and Kemper, Alfons and Neumann, Thomas},
  booktitle={Cidr},
  year={2017}
}

@inproceedings{wu2021unified,
  title={A unified deep model of learning from both data and queries for cardinality estimation},
  author={Wu, Peizhi and Cong, Gao},
  booktitle={Proceedings of the 2021 International Conference on Management of Data},
  pages={2009--2022},
  year={2021}
}

@article{zhu2020flat,
  title={FLAT: fast, lightweight and accurate method for cardinality estimation},
  author={Zhu, Rong and Wu, Ziniu and Han, Yuxing and Zeng, Kai and Pfadler, Andreas and Qian, Zhengping and Zhou, Jingren and Cui, Bin},
  journal={arXiv preprint arXiv:2011.09022},
  year={2020}
}

@inproceedings{10.1145/3514221.3517896,
author = {Hu, Xiao and Liu, Yuxi and Xiu, Haibo and Agarwal, Pankaj K. and Panigrahi, Debmalya and Roy, Sudeepa and Yang, Jun},
title = {Selectivity Functions of Range Queries are Learnable},
year = {2022},
isbn = {9781450392495},
booktitle = {SIGMOD},
pages = {959–972},
numpages = {14},
location = {Philadelphia, PA, USA},
series = {SIGMOD '22}
}

@inproceedings{li2022warper,
  title={Warper: Efficiently Adapting Learned Cardinality Estimators to Data and Workload Drifts},
  author={Li, Beibin and Lu, Yao and Kandula, Srikanth},
  booktitle={Proceedings of the 2022 International Conference on Management of Data},
  year={2022}
}

@article{10.14778/3583140.3583164,
    author = {Negi, Parimarjan and Wu, Ziniu and Kipf, Andreas and Tatbul, Nesime and Marcus, Ryan and Madden, Sam and Kraska, Tim and Alizadeh, Mohammad},
    title = {Robust Query Driven Cardinality Estimation under Changing Workloads},
    year = {2023},
    issue_date = {February 2023},
    publisher = {VLDB Endowment},
    volume = {16},
    number = {6},
    issn = {2150-8097},
    url = {https://doi.org/10.14778/3583140.3583164},
    doi = {10.14778/3583140.3583164},
    journal = {Proc. VLDB Endow.},
    month = {feb},
    pages = {1520–1533},
    numpages = {14}
}

@article{10.1145/3588713,
    author = {Kurmanji, Meghdad and Triantafillou, Peter},
    title = {Detect, Distill and Update: Learned DB Systems Facing Out of Distribution Data},
    year = {2023},
    issue_date = {May 2023},
    publisher = {Association for Computing Machinery},
    address = {New York, NY, USA},
    volume = {1},
    number = {1},
    url = {https://doi.org/10.1145/3588713},
    doi = {10.1145/3588713},
    journal = {Proc. ACM Manag. Data},
    month = {may},
    articleno = {33},
    numpages = {27}
}

@article{10.1145/3639293,
    author = {Wu, Peizhi and Ives, Zachary G.},
    title = {Modeling Shifting Workloads for Learned Database Systems},
    year = {2024},
    issue_date = {February 2024},
    publisher = {Association for Computing Machinery},
    address = {New York, NY, USA},
    volume = {2},
    number = {1},
    url = {https://doi.org/10.1145/3639293},
    doi = {10.1145/3639293},
    journal = {Proc. ACM Manag. Data},
    month = {mar},
    articleno = {38},
    numpages = {27}
}

@article{10.14778/3329772.3329780,
    author = {Dutt, Anshuman and Wang, Chi and Nazi, Azade and Kandula, Srikanth and Narasayya, Vivek and Chaudhuri, Surajit},
    title = {Selectivity estimation for range predicates using lightweight models},
    year = {2019},
    issue_date = {May 2019},
    publisher = {VLDB Endowment},
    volume = {12},
    number = {9},
    issn = {2150-8097},
    url = {https://doi.org/10.14778/3329772.3329780},
    doi = {10.14778/3329772.3329780},
    journal = {Proc. VLDB Endow.},
    month = may,
    pages = {1044–1057},
    numpages = {14}
}

@inproceedings{li2021cardinality,
    author = {Li, Beibin and Lu, Yao and Wang, Chi and Kandula, Srikanth},
    title = {Cardinality Estimation: Is Machine Learning a Silver Bullet?},
    booktitle = {AIDB},
    year = {2021},
    month = {August},
    url = {https://www.microsoft.com/en-us/research/publication/cardinality-estimation-is-machine-learning-a-silver-bullet/},
}

@article{10.14778/3476249.3476254,
    author = {Liu, Jie and Dong, Wenqian and Zhou, Qingqing and Li, Dong},
    title = {Fauce: fast and accurate deep ensembles with uncertainty for cardinality estimation},
    year = {2021},
    issue_date = {July 2021},
    publisher = {VLDB Endowment},
    volume = {14},
    number = {11},
    issn = {2150-8097},
    url = {https://doi.org/10.14778/3476249.3476254},
    doi = {10.14778/3476249.3476254},
    journal = {Proc. VLDB Endow.},
    month = jul,
    pages = {1950–1963},
    numpages = {14}
}

@misc{fullversion,
      title={[Full Version] SWatcher+: A Holistic Approach to Deploying Learned Cardinality Estimators in a Dynamic Setting}, 
      author={Anonymous Author(s)},
      year={2026},
      url={https://anonymous.4open.science/r/Watcher-SIGMOD2027/fullversion.pdf}, 
}

@misc{codes,
      title={[Codes] SWatcher+: A Holistic Approach to Deploying Learned Cardinality Estimators in a Dynamic Setting}, 
      author={Anonymous Author(s)},
      year={2026},
      url={https://anonymous.4open.science/r/Watcher-SIGMOD2027/README.md}, 
}

@article{jgmp,
    author = {Reiner, Silvan and Grossniklaus, Michael},
    title = {Sample-Efficient Cardinality Estimation Using Geometric Deep Learning},
    year = {2023},
    issue_date = {December 2023},
    publisher = {VLDB Endowment},
    volume = {17},
    number = {4},
    issn = {2150-8097},
    url = {https://doi.org/10.14778/3636218.3636229},
    doi = {10.14778/3636218.3636229},
    journal = {Proc. VLDB Endow.},
    month = dec,
    pages = {740–752},
    numpages = {13}
}

@misc{jgmp_code,
  author       = {Reiner, Silvan and Grossniklaus, Michael},
  title        = {Github repository for JGMP},
  howpublished = {\url{https://github.com/dbis-ukon/jgmp}},
  note         = {Accessed: March 27, 2025},
  year         = 2023
}

@article{10.14778/3476249.3476259,
author = {Negi, Parimarjan and Marcus, Ryan and Kipf, Andreas and Mao, Hongzi and Tatbul, Nesime and Kraska, Tim and Alizadeh, Mohammad},
title = {Flow-loss: learning cardinality estimates that matter},
year = {2021},
issue_date = {July 2021},
publisher = {VLDB Endowment},
volume = {14},
number = {11},
issn = {2150-8097},
url = {https://doi.org/10.14778/3476249.3476259},
doi = {10.14778/3476249.3476259},
journal = {Proc. VLDB Endow.},
month = jul,
pages = {2019–2032},
numpages = {14}
}

@inproceedings{DBLP:conf/edbt/MullerWL23,
  author       = {Magnus M{\"{u}}ller and
                  Lucas Woltmann and
                  Wolfgang Lehner},
  editor       = {Julia Stoyanovich and
                  Jens Teubner and
                  Nikos Mamoulis and
                  Evaggelia Pitoura and
                  Jan M{\"{u}}hlig and
                  Katja Hose and
                  Sourav S. Bhowmick and
                  Matteo Lissandrini},
  title        = {Enhanced Featurization of Queries with Mixed Combinations of Predicates
                  for ML-based Cardinality Estimation},
  booktitle    = {Proceedings 26th International Conference on Extending Database Technology,
                  {EDBT} 2023, Ioannina, Greece, March 28-31, 2023},
  pages        = {273--284},
  publisher    = {OpenProceedings.org},
  year         = {2023},
  url          = {https://doi.org/10.48786/edbt.2023.22},
  doi          = {10.48786/EDBT.2023.22},
  bibsource    = {dblp computer science bibliography, https://dblp.org}
}

@inproceedings{10.1145/3514221.3526154,
author = {Kim, Kyoungmin and Jung, Jisung and Seo, In and Han, Wook-Shin and Choi, Kangwoo and Chong, Jaehyok},
title = {Learned Cardinality Estimation: An In-depth Study},
year = {2022},
isbn = {9781450392495},
publisher = {Association for Computing Machinery},
address = {New York, NY, USA},
url = {https://doi.org/10.1145/3514221.3526154},
doi = {10.1145/3514221.3526154},
booktitle = {Proceedings of the 2022 International Conference on Management of Data},
pages = {1214–1227},
numpages = {14},
location = {Philadelphia, PA, USA},
series = {SIGMOD '22}
}

@article{10.1145/3654932,
    author = {Heddes, Mike and Nunes, Igor and Givargis, Tony and Nicolau, Alex},
    title = {Convolution and Cross-Correlation of Count Sketches Enables Fast Cardinality Estimation of Multi-Join Queries},
    year = {2024},
    issue_date = {June 2024},
    publisher = {Association for Computing Machinery},
    address = {New York, NY, USA},
    volume = {2},
    number = {3},
    url = {https://doi.org/10.1145/3654932},
    doi = {10.1145/3654932},
    journal = {Proc. ACM Manag. Data},
    month = may,
    articleno = {129},
    numpages = {26}
}

@inproceedings{LimeQO,
    author = {Yi, Zixuan and Tian, Yao and Ives, Zachary G. and Marcus, Ryan},
    title = {Low Rank Approximation for Learned Query Optimization},
    year = {2024},
    isbn = {9798400706806},
    publisher = {Association for Computing Machinery},
    address = {New York, NY, USA},
    url = {https://doi.org/10.1145/3663742.3663974},
    doi = {10.1145/3663742.3663974},
    booktitle = {Proceedings of the Seventh International Workshop on Exploiting Artificial Intelligence Techniques for Data Management},
    articleno = {4},
    numpages = {5},
    location = {Santiago, AA, Chile},
    series = {aiDM '24}
}

@article{BayesQO,
    author = {Tao, Jeffrey and Maus, Natalie and Jones, Haydn and Zeng, Yimeng and Gardner, Jacob R. and Marcus, Ryan},
    title = {Learned Offline Query Planning via Bayesian Optimization},
    year = {2025},
    issue_date = {June 2025},
    publisher = {Association for Computing Machinery},
    address = {New York, NY, USA},
    volume = {3},
    number = {3},
    url = {https://doi.org/10.1145/3725316},
    doi = {10.1145/3725316},
    journal = {Proc. ACM Manag. Data},
    month = jun,
    articleno = {179},
    numpages = {29}
}

@misc{stage,
      title={Stage: Query Execution Time Prediction in Amazon Redshift}, 
      author={Ziniu Wu and Ryan Marcus and Zhengchun Liu and Parimarjan Negi and Vikram Nathan and Pascal Pfeil and Gaurav Saxena and Mohammad Rahman and Balakrishnan Narayanaswamy and Tim Kraska},
      year={2024},
      eprint={2403.02286},
      archivePrefix={arXiv},
      primaryClass={cs.DB},
      url={https://arxiv.org/abs/2403.02286}, 
}

@inproceedings{LEO,
    author = {Stillger, Michael and Lohman, Guy M. and Markl, Volker and Kandil, Mokhtar},
    title = {LEO - DB2's LEarning Optimizer},
    year = {2001},
    isbn = {1558608044},
    publisher = {Morgan Kaufmann Publishers Inc.},
    address = {San Francisco, CA, USA},
    booktitle = {Proceedings of the 27th International Conference on Very Large Data Bases},
    pages = {19–28},
    numpages = {10},
    series = {VLDB '01}
}

@misc{sql_server_ce_feedback,
  author       = {Microsoft SQL Server},
  title        = {Cardinality estimation (CE) feedback},
  howpublished = {\url{https://learn.microsoft.com/en-us/sql/relational-databases/performance/intelligent-query-processing-cardinality-estimation-feedback?view=sql-server-ver16}},
  note         = {Accessed: Sep 18, 2025},
  year         = 2025
}

@inproceedings{Bao,
    author = {Marcus, Ryan and Negi, Parimarjan and Mao, Hongzi and Tatbul, Nesime and Alizadeh, Mohammad and Kraska, Tim},
    title = {Bao: Making Learned Query Optimization Practical},
    year = {2021},
    isbn = {9781450383431},
    publisher = {Association for Computing Machinery},
    address = {New York, NY, USA},
    url = {https://doi.org/10.1145/3448016.3452838},
    doi = {10.1145/3448016.3452838},
    booktitle = {Proceedings of the 2021 International Conference on Management of Data},
    pages = {1275–1288},
    numpages = {14},
    location = {Virtual Event, China},
    series = {SIGMOD '21}
}

@inproceedings{Balsa,
    author = {Yang, Zongheng and Chiang, Wei-Lin and Luan, Sifei and Mittal, Gautam and Luo, Michael and Stoica, Ion},
    year = {2022},
    month = {06},
    pages = {931-944},
    title = {Balsa: Learning a Query Optimizer Without Expert Demonstrations},
    doi = {10.1145/3514221.3517885}
}

@article{Lero,
    author = {Zhu, Rong and Chen, Wei and Ding, Bolin and Chen, Xingguang and Pfadler, Andreas and Wu, Ziniu and Zhou, Jingren},
    title = {Lero: A Learning-to-Rank Query Optimizer},
    year = {2023},
    issue_date = {February 2023},
    publisher = {VLDB Endowment},
    volume = {16},
    number = {6},
    issn = {2150-8097},
    url = {https://doi.org/10.14778/3583140.3583160},
    doi = {10.14778/3583140.3583160},
    journal = {Proc. VLDB Endow.},
    month = feb,
    pages = {1466–1479},
    numpages = {14}
}

@article{wu2025practicaltheorygeneralizationselectivity,
author = {Wu, Peizhi and Xu, Haoshu and Marcus, Ryan and Ives, Zachary G.},
title = {A Practical Theory of Generalization in Selectivity Learning},
year = {2025},
issue_date = {February 2025},
publisher = {VLDB Endowment},
volume = {18},
number = {6},
issn = {2150-8097},
url = {https://doi.org/10.14778/3725688.3725708},
doi = {10.14778/3725688.3725708},
journal = {Proc. VLDB Endow.},
month = aug,
pages = {1811–1824},
numpages = {14}
}

@misc{naru_codebase,
  title = {Codebase of Naru},
  author = {Yang, Z. and Liang, E. and Kamsetty, A. and Wu, C. and Duan, Y. and Chen, X. and Abbeel, P. and Hellerstein, J. M. and Krishnan, S. and Stoica, I.},
  year = {2019},
  howpublished = {\url{https://github.com/naru-project/naru/tree/master}},
  note = {Accessed: 2026-04-13}
}

@misc{job_light_sql,
  title = {Job-light Workload},
  author = {Andreas Kipf and
                  Thomas Kipf and
                  Bernhard Radke and
                  Viktor Leis and
                  Peter A. Boncz and
                  Alfons Kemper},
  year = {2019},
  howpublished = {\url{https://github.com/andreaskipf/learnedcardinalities/blob/master/workloads/job-light.sql}},
  note = {Accessed: 2026-05-11}
}

@misc{imdb,
  title = {IMDB},
  author = {IMDB.com, Inc},
  year = {2019},
  howpublished = {\url{https://www.imdb.com/}},
  note = {Accessed: 2026-05-11}
}

@misc{deepdb_codebase,
  title = {DeepDB codebase},
  author = {Hilprecht, B. and Schmidt, A. and Kulessa, M. and Molina, A. and Kersting, K. and Binnig, C.},
  year = {2019},
  howpublished = {\url{https://github.com/DataManagementLab/deepdb-public/tree/master}},
  note = {Accessed: 2026-04-13}
}

@misc{postgre_wal,
  title = {PostgreSQL: Write-Ahead Logging (WAL)},
  author = {The PostgreSQL Global Development Group},
  year = {2026},
  howpublished = {\url{https://www.postgresql.org/docs/current/wal-intro.html}},
  note = {Accessed: 2026-05-11}
}



\end{document}